\documentclass[aps,prb,preprint,groupedaddress,floatfix]{revtex4}
\usepackage{graphicx,amssymb}
\begin{document}
\title{Evaluation of the grand-canonical partition function using Expanded Wang-Landau simulations. V. Impact of an electric field on the thermodynamic properties and ideality contours of water.}
\author{Caroline Desgranges and Jerome Delhommelle}
\affiliation{Department of Chemistry, 151 Cornell Street Stop 9024, University of North Dakota, Grand Forks ND 58202}

\date{\today}

\begin{abstract}
Using molecular simulation, we assess the impact of an electric field on the properties of water, modeled with the SPC/E potential, over a wide range of states and conditions. Electric fields of the order of $0.1~V/$~\AA~and beyond are found to have a significant impact on the grand-canonical partition function of water, resulting in shifts in the chemical potential at the vapor-liquid coexistence of up to $20$~\%. This, in turn, leads to increases in the critical temperatures by close to $7$~\% for a field of $0.2~V/$\AA, to lower vapor pressures, and to much larger entropies of vaporization (by up to $35$~\%). We interpret these results in terms of the greater density change at the transition and of the increased structural order resulting from the applied field. The thermodynamics of compressed liquids and of supercritical water are also analyzed over a wide range of pressures, leading to the determination of the Zeno line and of the curve of ideal enthalpy that span the supercritical region of the phase diagram. Rescaling the phase diagrams obtained for the different field strength by their respective critical properties allows us to draw a correspondence between these systems for fields of up to $0.2~V/$\AA.
\end{abstract}

\maketitle

\section{Introduction}
The properties of water subjected to an electric field are of key importance for many applications, ranging from nanofluidics~\cite{nano1,nano2,nano3,nano4}, biochemistry~\cite{bio1,bio2} and chemical processes~\cite{dist1,dist2}. This is, for instance, the case in living organisms, where electric fields control the orientation of the water molecules and thus the proton transfer direction driving ATP synthesis~\cite{bio2}. Similarly, in atmospheric sciences, ions are well known to have a direct impact on the nucleation process by, depending upon the conditions, promoting or preventing the transition from vapor to liquid through the nucleation of droplets~\cite{atm1,atm2,atm3}. Alternatively, an electric field can also be used as a great tool to control the behavior, trigger phase transitions and tune the properties of liquids and materials~\cite{pgku,pol1,pol2,pol3}. 

However, the response of a polar liquid subjected to very strong fields still remains poorly understood. Recent work show that the critical temperature of water subjected to very large electric fields is lower than that of the bulk~\cite{down1,down2}, while others indicate either the opposite behavior~\cite{leibler,Gabor,ZengField,SiepmannField} or, in the case of nanoconfined water, a complex dependence of the critical temperature on the extent of the nanoconfinement and the field strength~\cite{ConfSingh}. There are also few studies available on the impact of the electric field on the Gibbs free energy and entropy of water. Furthermore, in recent years, the elucidation of the thermodynamics of supercritical water has drawn considerable interest in geochemistry~\cite{kessel2005water} and as a green solvent for the catalytic conversion of biomass into fuel~\cite{peterson2008thermochemical}, and it is unknown how strong fields change the thermodynamics and structure of water in the supercritical region of the phase diagram. The goal of this work is twofold: (i) we extend the recently developed Expanded Wang-Landau (EWL) simulation method~\cite{PartI,PartII,PartIII,PartIV} to determine the thermodynamic properties of systems subjected to an external field and (ii) we apply the resulting method to analyze the impact of the electric field on the properties of water over a wide range of conditions, i.e. at the vapor-liquid phase boundary, for compressed liquids as well as under supercritical conditions. In particular, in the supercritical region of the phase diagram, we focus on analyzing the effect of the field on the ideality contours~\cite{nedostup2013asymptotic,apfelbaum2013regarding,kutney2000zeno,wei2013isomorphism,Leanna,Landon,Apfelbaum2016}, known as the Zeno line and the curve of ideal enthalpy, to establish a correspondence between the results obtained for different fields. These ideality contours have recently emerged as a new way to bridge the gap in our understanding of supercritical fluids~\cite{brazhkin2013liquid,brazhkin2011van,Morita,brazhkin2012supercritical} and have paved the way for the development of new similarity laws~\cite{apfelbaum2013regarding} and maps of the supercritical region of the phase diagram~\cite{apfelbaum2015similarity,Abigail}. 

This paper is organized as follows. In the next section, we discuss how we extend the EWL method to a polar fluid subjected to an electric field. We then detail the simulation model as well as the technical details before presenting the results obtained in this work. We start by discussing how the electric field impacts the grand-canonical partition function for water and from there, examine the effect of the field on the thermodynamics of phase transition, on the location of the critical point and on the response of compressed liquids of water and supercritical water to the field. We particularly focus on the interplay between the changes in density and structure and the thermodynamic properties of water under an electric field. We finally discuss the impact of the field on the ideality contours of water before drawing the main conclusions of this work in the last section.

\section{Expanded Wang-Landau sampling for systems subjected to an external field}

\subsection{Theoretical framework}

The goal of the recently developed Expanded Wang-Landau (EWL) simulation method consists in determining a high-accuracy estimate for the grand-canonical partition function, which, in turn, yields all thermodynamic properties of the system through the statistical mechanics formalism~\cite{McQuarrie}. The first papers of the series have discussed developments of the method as applied to single-component systems~\cite{PartI,PartII} and mixtures~\cite{PartIII} modeled with classical and quantum (tight-binding)~\cite{PartIV} force fields. Here we extend this approach to the case of systems subjected to an external field and consider the case of water subjected to an electric field. The grand-canonical partition function~\cite{McQuarrie} is given by
 \begin{equation}
\Theta(\mu,V,T)= \sum_{N=0}^\infty Q(N,V,T) \exp (\beta \mu N)
\label{ThetamuVT}   
\end{equation}
where $\beta=1/k_BT$, $N$ the number of molecules, $\mu$ the chemical potential and $Q(N,V,T)$ is the canonical partition function written as
 \begin{equation}
Q(N,V,T)= {(q_{trans} q_{int})^N \over N!}  \int \exp\left(-\beta U(\mathbf{\Gamma})\right)d{\mathbf {\Gamma}}
\label{Q_NVT}   
\end{equation}
where $q_{trans}=[{2\pi M k_B T \over h^2}]^{3/2}V= {V \over \Lambda^3}$ is the translational partition function for an individual water molecule of mass $M$, $q_{int}$ is the intramolecular partition function and $\mathbf {\Gamma}$ denotes a specific configuration of the system. The equation for $q_{int}$ depends on the assumptions underlying the force field used for water as discussed in the next section. 

In Eq.~\ref{Q_NVT}, $U(\mathbf{\Gamma})$ denotes the potential energy of the system for a given configuration. $U$ is calculated as the sum of the interaction energy between water molecules and of the energy resulting from the interaction with the electric field. For a system of $N$ molecules, the energy due to the field is obtained through
 \begin{equation}
U_f=-\sum_{i=1}^N {\mathbf {m_i . E}}
\label{UF}
\end{equation}
where ${\mathbf {m_i}}$ is the dipole moment of water molecule $i$ and ${\mathbf {E}}$ is the applied electric field.

EWL simulations take advantage of an efficient scheme for the insertion/deletion of molecules, known as the expanded ensemble approach~\cite{Lyubartsev,expanded,Paul,Escobedo,MV2,Singh,Shi,Jason,Aaron,Erica,Andrew}, which splits the insertion and deletion of a full molecule into $M$ stages. The implementation of efficient schemes for the insertions/deletion steps, such as e.g. in expanded ensemble-transition matrix Monte Carlo methods~\cite{Rane} or with the continuous fraction component methods~\cite{Yee,Sikora}, has been shown to yield very accurate results. Here, the combination of a Wang-Landau sampling with the expanded ensemble approach ensures an efficient sampling of all possible $N$ values and results in highly accurate predictions for the thermodynamic properties in the low temperature-high density regime~\cite{PartI,Camp}. In the EWL method, the simulated system is composed of $N$ full molecules and of a fractional molecule at stage $l$ (with $0 \le l \le M-1$). The insertion/deletions steps are handled through changes in the value of $l$. If, during the simulation, $l$ is increased beyond $M$, the fractional molecule becomes a full molecule and a new fractional molecule at stage $l-M$ is created. This results in the insertion of a new full molecule as the system now contains $N+1$ full molecule and a new fractional atom at stage $l-M$. The deletion of a molecule is similarly achieved though a decrease of the value of $l$ for the fractional molecule. Finally, when $l=0$, the fractional molecule is void and the system contains $N$ full molecules. The resulting simplified expanded grand-canonical ensemble (SEGC) partition function~\cite{PartI} for this system is
\begin{equation}
\Theta_{SEGC}(\mu,V,T)= \sum_{N=0}^\infty \sum_{l=0}^{M-1} Q(N,V,T,l) \exp (\beta \mu N) \\
\label{SEGC}   
\end{equation}
in which $Q(N,V,T,l)$ is the canonical partition function for a system of $N$ full atoms and a fractional atom at stage $l>0$, given by
 \begin{equation}
Q(N,V,T,l)=  {q_{trans}^N q_{int}^N q_{l,trans} q_{l,int} \over N!} \int \exp\left(-\beta U({\mathbf {\Gamma}})\right) d{\mathbf {\Gamma}}
\label{Q_NVTfrac}  
\end{equation}
Here, the mass, as well as the moments of inertia, for the fractional molecule are chosen to be the same as that of a full molecule, leading to identical translational and rotational partition functions, $q_{l,trans}=q_{trans}$ and $q_{l,rot}=q_{rot}$. 

The $Q(N,V,T,l)$ functions are then determined numerically during the EWL simulations, through the iterative evaluation of the biased distribution $p_{bias}(\mathbf {\Gamma}, N, l)$~\cite{PartI,PartII,PartIII,PartIV}. For Wang-Landau simulations~\cite{Wang1,Wang2,Shell,Shell2,Shell3,dePablo,Rampf,Camp,WLHMC,Copper,MolPhys,KennethI,Malakis}, the Metropolis criterion for a move from an old state ($\mathbf {\Gamma_o},N_o,l_o$) to a new state ($\mathbf {\Gamma_n},N_n,l_n$) is given by
 \begin{equation}
acc(o \to n)=min\left[ 1, {p_{bias}(\mathbf {\Gamma_n}, N_n, l_n) \over p_{bias}(\mathbf {\Gamma_o}, N_o, l_o)} \right]
\label{Metro}   
\end{equation}

Following the derivation for $p_{bias}$ for the EWL simulations~\cite{PartI,PartII,PartIII,PartIV}, we obtain for the case of water the following equation for $p_{bias}$
\begin{equation}
{p_{bias}(\mathbf {\Gamma}, N, l)} = { \exp\left( - \beta \left [U(\Gamma) -\mu N \right] \right) \over { N! q_{trans}^{3(N+1)} q_{int}^{3(N+1)} Q(N,V,T,l)}}\\
\label{pbias}   
\end{equation}

This leads to the numerical determination of $p_{bias}$ and thus of $Q(N,V,T,l)$. A specific advantage of carrying out a Wang-Landau sampling in the grand-canonical ensemble is that the variable sampled, i.e. $N$ the number of molecules, is a discrete function, which circumvents the discretization of the energy range that would be required in other ensembles~\cite{Do}, and allows the simulations to obtain accurate free energy measurements. The grand-canonical partition function $\Theta(\mu,V,T)$ can then be calculated for any value of $\mu$ from Eq.~\ref{ThetamuVT} using the numerical values of $Q(N,V,T,l=0)$ (we note this function as $Q(N,V,T)$ from now on and drop the $l=0$ specification). The conditions for vapor-liquid coexistence are determined from the number distribution $p(N)$ as follows. We start from $p(N)$ defined as
\begin{equation}
p(N) = {Q(N,V,T)  \exp\left(\beta \mu N\right) \over \Theta(\mu,V,T)}\\
\label{pN}   
\end{equation}
We then solve numerically the equation below to determine the chemical potential at coexistence
\begin{equation}
\sum_{N=0}^{N_b} p(N) = \sum_{N_b}^{\infty} p(N)
\label{coex}
\end{equation} 
where $N_b$ is the point at which the function $p(N)$ reaches its minimum, and the left hand side and the right hand side of the equation correspond to the probability of the vapor and of the liquid phase, respectively. The other thermodynamic properties can then be determined through the usual statistical mechanics relations~\cite{McQuarrie,PartI}. 

\subsection{Simulation Models}
We use the SPC/E force field~\cite{berendsen1987missing} to model $H_2O$. Each molecule is described as a distribution of three LJ sites and three point charges (one on each atom) with the interaction between atoms $(i,j)$ as
\begin{equation}
\phi(r_{ij})=4\epsilon_{ij} \left[ {\left({\sigma_{ij} \over r_{ij}}\right) ^{12}- \left({\sigma_{ij} \over r_{ij}}\right) ^6} \right] + {q_i q_j \over {4 \pi \epsilon_0 r_{ij}}}
\label{TraPPE}   
\end{equation}
The parameters for the SPC/E model are taken from Berendsen {\it et al}~\cite{berendsen1987missing}. To model the interaction between an atom of the fractional molecule with an atom of a full molecule we scale the interaction parameters $\epsilon_{ij},\sigma_{ij}$ and the product $q_i q_j$ by $(l/M)^{1/3}$, $(l/M)^{1/4}$ and $(l/M)^{1/3}$, respectively. We scale the size of the fractional molecule (i.e. the bond length between $H$ and $O$) by $(l/M)^{1/4}$. 

The SPC/E model~\cite{berendsen1987missing} treats water as a rigid molecule. This implies that the effect of the intramolecular vibrations on the thermodynamics of water are not taken into account, and that, in this case, $q_{int}$ is equal to the rotational partition function of the water molecule. This gives
 \begin{equation}
q_{int}={{\sqrt \pi}\over \sigma} {\left( 8 \pi^2 I_A k_B T \over h^2\right)^{1/2} }{\left( 8 \pi^2 I_B k_B T \over h^2\right)^{1/2} }{\left( 8 \pi^2 I_C k_B T \over h^2\right)^{1/2} }
\label{qint}   
\end{equation}
where $\sigma$ is the symmetry number (equal to $2$ in the case of water), and $I_A$, $I_B$ and $I_C$ are the 3 principal moments of inertia.

The electric field $\mathbf E$ is applied along the $x$-axis. This means that the interaction between the dipole moment $\mathbf m_i$ of water molecule $i$ can be calculated as follows:
\begin{equation}
U_f=- \sum_{i=1}^N \sum_{\alpha=1}^3 q_{\alpha} x_{i,\alpha} E
\end{equation}
where $E$ is the norm of the field $\mathbf E$, $q_{\alpha}$ is the point charge associated with atom $\alpha$ of molecule $i$ and $x_{i,\alpha}$ is the coordinates of atom $\alpha$ of molecule $i$ along the $x$-axis.

The use of a classical force field, like the SPC/E potential, to model water under an electric field~\cite{ZengField,SiepmannField,aragones2011phase} implies that the dissociation of water molecules is not taken into account in this work. Recent {\it ab initio} molecular dynamics simulations (MD)~\cite{saitta2012ab} have identified a dissociation threshold for water molecules for an electric field of about $0.35$~V/\AA. This is in agreement with the findings from previous experimental work~\cite{stuve2012ionization,rothfuss2003influence}, that reported the onset of water dissociation under external fields of about $0.32$~V/\AA~to $0.44$~V/\AA, and from previous simulation results~\cite{geissler2001autoionization}, that showed that external fields of $0.3$~V/\AA~to $0.6$~V/\AA~enhance water ionization. In this work, we consider fields $E$ of $0.05$~V/\AA, $0.1$~V/\AA and $0.2$~V/\AA, i.e. well below the dissociation threshold and for which the dissociation of water molecules are extremely rare and short-lived~\cite{saitta2012ab}. We also include, for comparison, results for a larger field of $0.5$~V/\AA, a field for which {\it ab initio} MD indicates that $8$~\% of water molecules are ionized~\cite{saitta2012ab}. Very interestingly, these results show that such a strong field yields to a qualitatively different behavior of water, even if water dissociation is not taken into account. Finally, we add that the use of a rigid model amounts to neglecting the elongation of the intramolecular $O-H$ bonds as a result of the electric field. However, according to the {\it ab initio} MD results~\cite{saitta2012ab}, the range of electric fields studied here as a very moderate impact on the $O-H$ bondlength. Saitta {\it et al.}'s work reveal a non-monotonic dependence of the bondlength on the electric field, with an $O-H$ bondlength varying between $1$~\AA~and $1.01$~\AA.

\subsection{Technical details}
We use 3 different types of Monte Carlo (MC) moves during the EWL simulations of water under an electric field. The first type of MC move ($37.5$\% of the total number of moves) correspond to the translation of a single water molecule (randomly chosen among the $N+1$ molecules, i.e. the $N$ full water molecules plus the fractional molecule). The second type of MC move ($37.5$\% of the total number of moves) involves the rotation of a single molecule (randomly selected as one of the $N$ full molecules or the fractional molecule). The $25$\% remaining moves are changes in $(N,l)$ values for the system, resulting in the sampling of the range of $N$, and hence densities, studied in this work. Here we carry out simulations of from $N=0$ to up to $N=300$ molecules in cubic cells with box lengths of $L=20$~\AA~for all systems, with the usual periodic boundary conditions~\cite{Allen}. For the $LJ$ part of the potential, we use a cutoff distance set to half the box length and apply tail corrections beyond that cutoff distance~\cite{Allen}. The long-range electrostatic interactions are handled using Ewald sums, with the screening parameter for the charge Gaussian distribution set to $5.6/L$ and the reciprocal cutoff vector set to $k_{max}=6(2\pi)/L$. The parameters for the $EWL$ simulations are the same as in prior work~\cite{PartIII}. The number of stages $M$ is set to $100$, the starting value for the convergence factor $f$ in the iterative Wang-Landau scheme to $e$, its final value to $10^{-8}$, with each $(N,l)$ value being visited at least 1000 times for a given value of $f$.

\section{Results and Discussion}

\subsection{Partition functions for water subjected to an electric field}

\begin{figure}
\begin{center}
\includegraphics*[width=8cm]{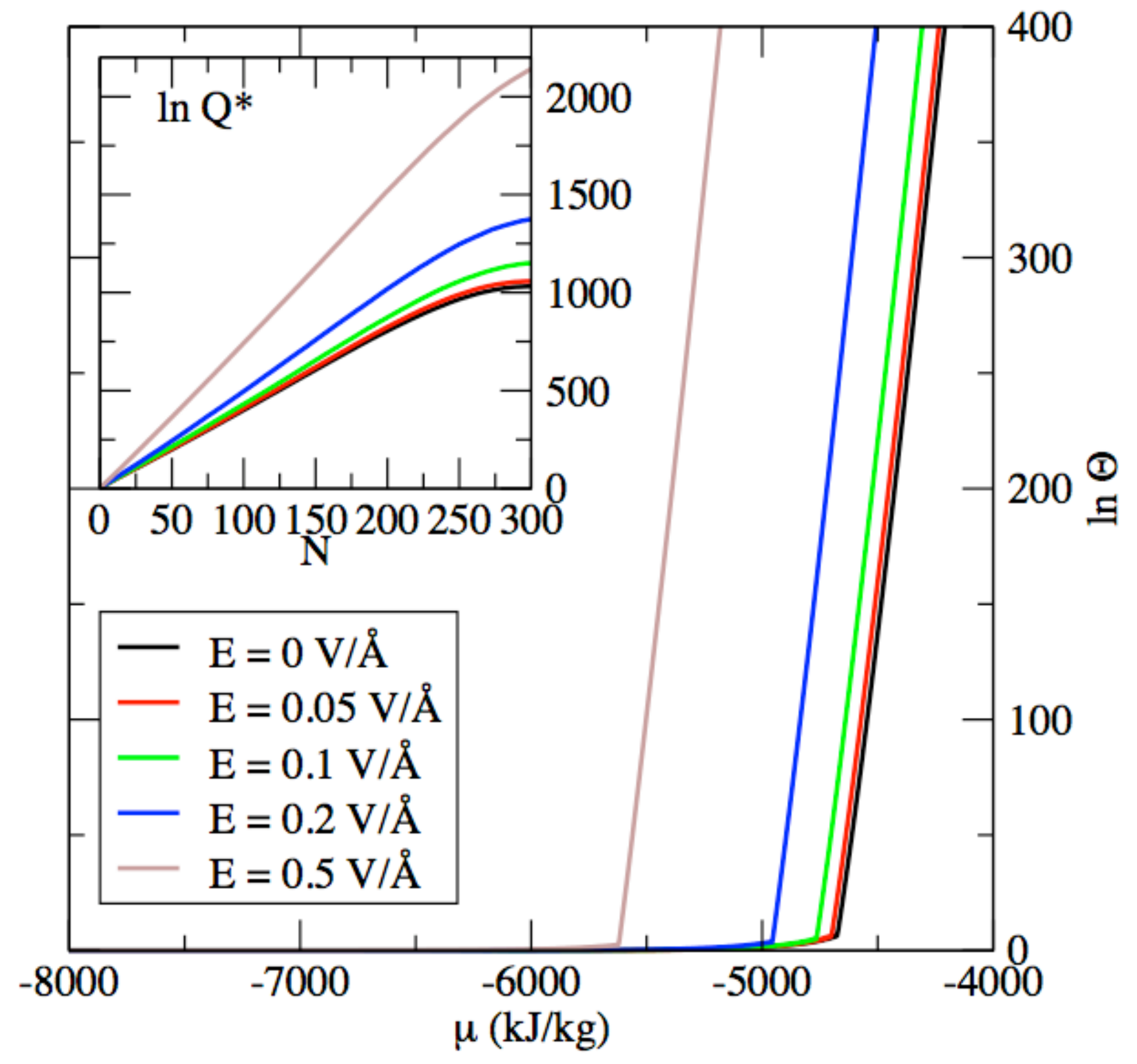}
\end{center}
\caption{Water at $T=575$~K: Logarithm of the grand-canonical partition function $\Theta (\mu,V,T)$ in the absence of field (black), for $E=0.05~V/$\AA, for $E=0.1~V/$\AA, for $E=0.2~V/$\AA~and for $E=0.5~V/$\AA~with as an inset, logarithm of the functions $Q^*(N,V,T)=Q(N,V,T)/q_{int}^N$, where $q_{int}$ is the internal partition function for a water molecule, as a function of the number of water molecules $N$ for different field strengths.}
\label{Fig1}
\end{figure}

We start by analyzing the results obtained for the grand-canonical partition function for water subjected to an electric field. Fig.~\ref{Fig1} shows that the behavior for this function at $T=575~K$ for increasing values of the field. We also include in Fig.~\ref{Fig1} the results obtained in the absence of field for comparison purposes. Fig.~\ref{Fig1} shows that increasing the field strength has two main effects on $\ln \Theta (\mu,V,T)$. First, it leads to a shift in the chemical potential for which $\ln \Theta$ exhibits a steep increase. Since this sharp increase corresponds to the vapor $\to$ liquid transition~\cite{jctc2015}, this means that $\mu_{coex}$, the chemical potential at coexistence, becomes lower and lower with the field. At $T=575~K$, and taking as reference $\mu_{coex}$ in the absence of field, applying a field of $0.05~V/$\AA~results in a decrease of $0.7$\% in $\mu_{coex}$, while a field of $0.1~V/$\AA~yields a decrease in $\mu_{coex}$ of $1.6~$\%. This decrease becomes even more significant as the field is further increased with a value of $\mu_{coex}$ lower by $6.3~$\% for a field of $0.2~V/$\AA~ and by $25~$\% for a field of $0.5~V/$\AA. This result can be directly connected to the increase in the $\ln Q(N,V,T)$ function, shown in the inset of Fig.~\ref{Fig1}, since the slope of $\ln Q(N,V,T)$ is proportional to the chemical potential. Second, the steep increase in $\ln \Theta (\mu,V,T)$ becomes sharper and sharper with the field strength. This implies that the difference between the vapor and liquid phases becomes more and more important as the field increases. A corollary to this result is the fact that for a given temperature, applying an electric field will result in making the two phases more and more different or, in other words, postponing (in terms of temperature) the onset of criticality, for which the difference between the properties of the two phases vanishes. This suggests that the critical point increases as a result of the increase in field strength. This point will be further studied in the next section.

\begin{figure}
\begin{center}
\includegraphics*[width=8cm]{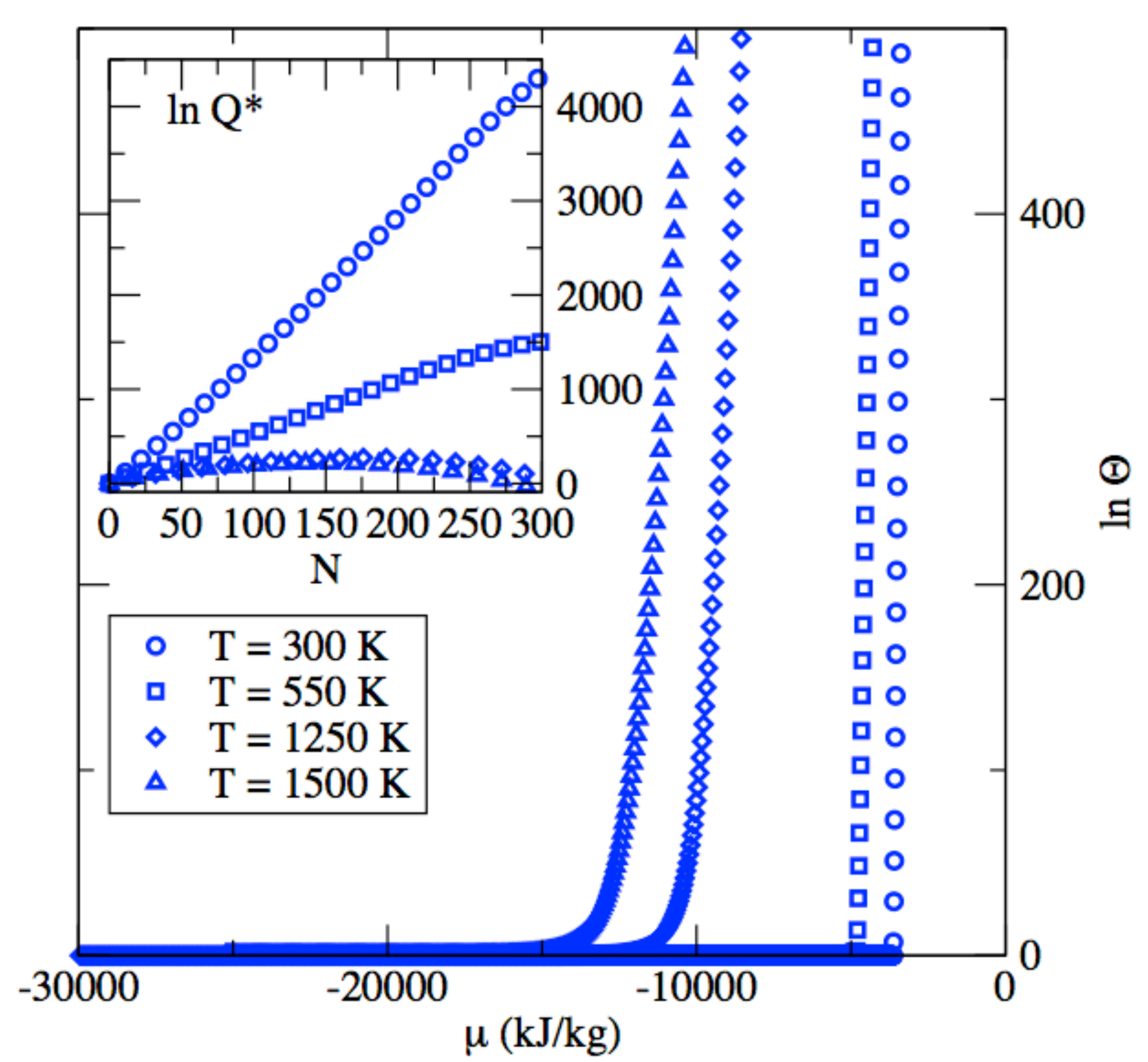}
\end{center}
\caption{Water subjected to $E=0.2~V/$\AA: temperature dependence of $\ln \Theta (\mu,V,T)$ and $\ln Q(N,V,T)$.}
\label{Fig2}
\end{figure}

We now focus on the effect of temperature for a fixed field strength. Fig.~\ref{Fig2} shows the temperature dependence of $\ln \Theta (\mu,V,T)$ for a field of $0.2~V/$\AA. The plots exhibit two qualitatively different behaviors. A sharp increase, corresponding to the fluid undergoing the vapor $\to$ liquid transition, is observed for the two lower temperatures. On the other hand, a much smoother increase in $\ln \Theta (\mu,V,T)$, typical of a supercritical fluid for which there is no longer a transition, is found for the two higher temperatures. Increasing the temperature has a direct effect on  $\mu_{coex}$ as shown by the shift observed between the results for $T=300~K$ and for $T=550~K$. At $T=550~K$, $\mu_{coex}$ is $38~$\% lower than $\mu_{coex}$ at $T=300~K$. As in Fig.~\ref{Fig1}, this shift in $\mu_{coex}$ can be correlated with the variations for the slope of $\ln Q(N,VT)$, shown in the inset of Fig.~\ref{Fig2}, which is found to decrease as the temperature increases from $300~K$ to $550~K$. We also find a markedly different behavior in $\ln Q(N,V,T)$ between the results for subcritical and supercritical fluids. While $\ln Q(N,V,T)$ is a monotonic function over the whole range of densities for subcritical fluids, we observe that, for a supercritical fluid, $\ln Q(N,V,T)$ is non-monotonic anymore and exhibits a maximum. Furthermore, $\ln Q(N,V,T)$ becomes less and less dependent on the temperature in the supercritical regime, which can be attributed to the lesser impact of the intermolecular interactions at very high temperatures.

\subsection{Thermodynamics of the vapor-liquid transition and critical properties}

\begin{figure}
\begin{center}
\includegraphics*[width=8cm]{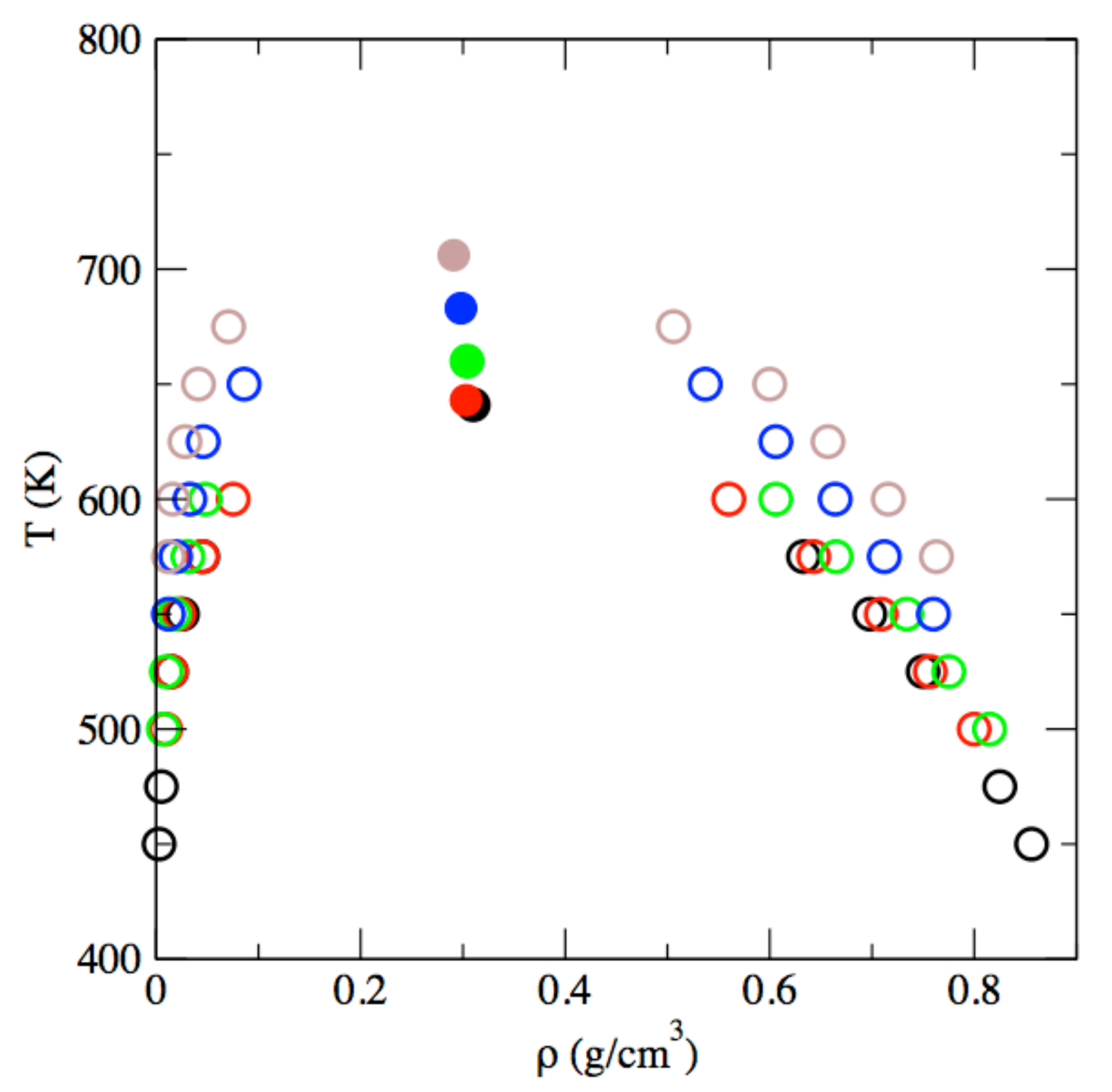}
\end{center}
\caption{Vapor-liquid equilibrium properties for water:  EWL densities at coexistence in the absence of field, for $E=0.05~V/$\AA, for $E=0.1~V/$\AA, for $E=0.2~V/$\AA~and for $E=0.5~V/$\AA~(same legend as in Fig.~\ref{Fig1}).}
\label{Fig3}
\end{figure}

Using the results obtained for the partition function, we calculate the number probability distribution $p(N)$ to determine the densities at coexistence for the vapor-liquid transition~\cite{PartI} using Eqs.~\ref{pN} and~\ref{coex}. We show in Fig.~\ref{Fig3}  the densities at coexistence obtained in the absence of field and for field strengths ranging from $0.05~V/$\AA~ to $0.5~V/$\AA. We also plot in Fig.~\ref{Fig3} the critical point for each field value. The critical parameters are obtained as follows. We use a scaling law for the critical temperature
\begin{equation}
\rho_l-\rho_v=B (T-T_c)^\beta
\label{scaleT}
\end{equation}
where B is a fitting parameter, $\beta$ is the 3D Ising critical exponent adjusted for real substances ($\beta=0.326$), and $\rho_l$ and $\rho_v$ are the densities for the liquid and vapor phases at coexistence given by the EWL simulations.
The critical density is calculated from the law of rectilinear diameters
\begin{equation}
{{\rho_l + \rho_v} \over 2} = \rho_c +A (T-T_c)
\end{equation}
where $\rho_c$ and A are two fitting parameters and $T_c$ is the estimate for the critical temperature obtained from Eq.~\ref{scaleT}. The critical points so obtained are plotted in Fig.~\ref{Fig3} and their numerical values are given in Table~\ref{param}.

The electric field has two main effects on the phase diagram of water. First, the whole phase diagram is shifted towards the higher temperatures. This can best be seen through the increase in $T_c$ which starts from $641~K$ in the absence of field, and increases by $0.3$\% for $E=0.05~V/$\AA~and by $3~$\% for $E=0.1~V/$\AA. This increase becomes even more pronounced as the field further increases with a $T_c$ for $E=0.2~V/$\AA~greater by $6.6~$\% than in the absence of field, and a $T_c$ for $E=0.5~V/$\AA~that is $10.1~$\% above the value of $T_c$ in the absence of field. Second, applying an electric field results in tilting the phase diagram, with a decrease in the critical density for very strong fields. More specifically, taking as the reference the critical density in the absence of field, we find that $\rho_c$ decreases by $1.9~$\% for $E=0.05~V/$\AA~, by $2.3~$\% for $E=0.1~V/$\AA~, by $3.9~$\% for $E=0.2~V/$\AA~ and by $6.1~$\% for $E=0.5~V/$\AA. This combined increase in $T_c$ and decrease in $\rho_c$ is consistent with the findings from Gibbs Ensemble Monte Carlo simulations on polar compounds~\cite{SiepmannField}. Comparing the results obtained for $T=575~K$, we find that the density of the liquid at coexistence increases with the field, whereas the density of the vapor at coexistence decreases with the field. Overall, the results confirm that fields of up to $0.05~V\/$\AA~have a limited impact on the densities at coexistence~\cite{ZengField} and on the critical parameters. This is consistent with the moderate changes in the partition functions for such fields, as shown in Fig.~\ref{Fig1} and discussed in Section A. We therefore focus in the rest of the paper on fields of the order of $0.1~V/$\AA~and above. 

\begin{figure}
\begin{center}
\includegraphics*[width=8cm]{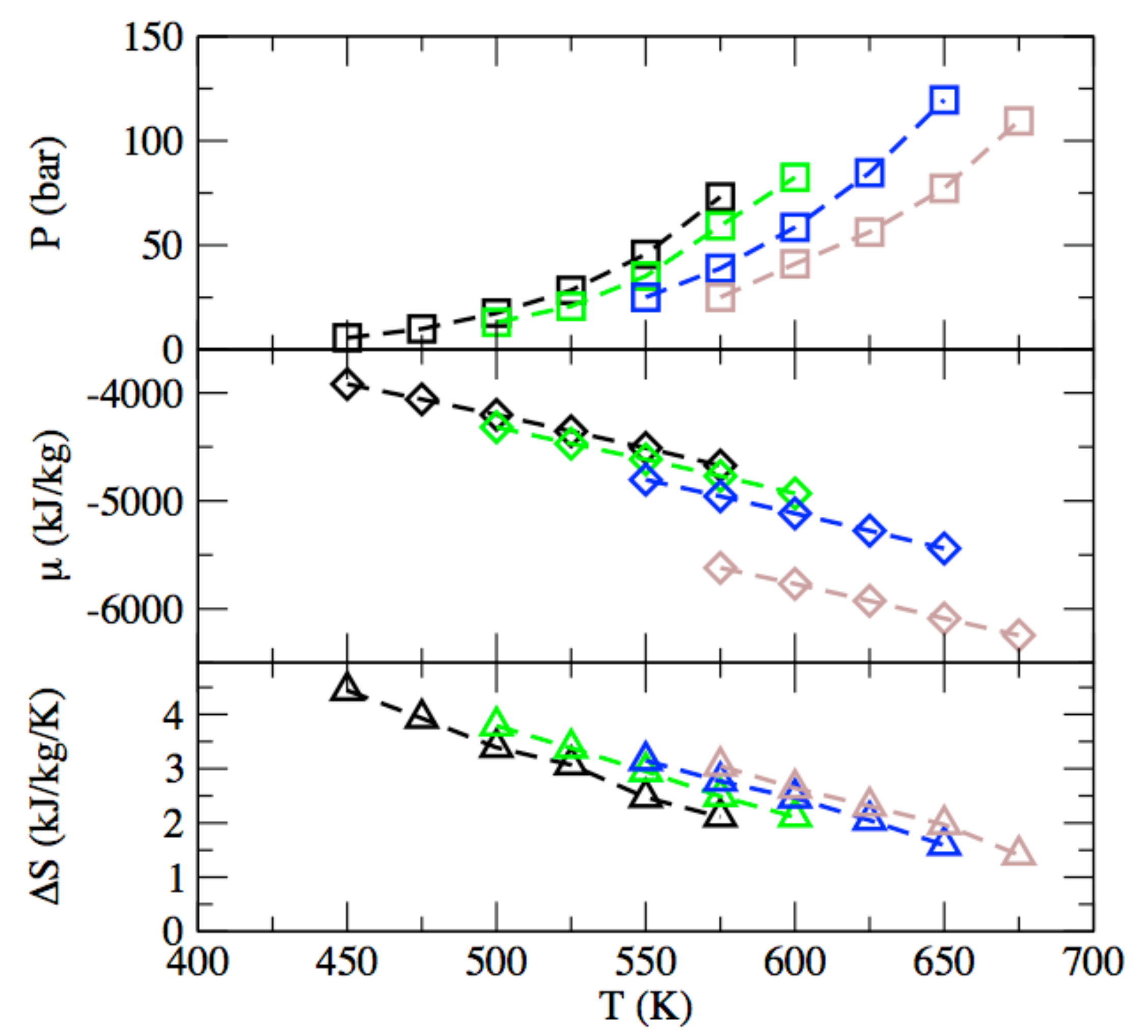}
\end{center}
\caption{Vapor-liquid equilibrium properties for water:  (Top) Pressure $P$ vs. temperature $T$ in the absence of field (black) and for field strengths of $E=0.1~V/$\AA~(green), $E=0.2~V/$\AA~(blue) and $E=0.5~V/$\AA~(brown), (Middle) Chemical potential $\mu$ at coexistence vs. $T$, and (Bottom) Entropy change $\Delta S$ vs. $T$.}
\label{Fig4}
\end{figure}

We now turn to the results obtained for the thermodynamic properties at coexistence. Fig.~\ref{Fig4} shows the vapor pressure obtained from the grand-canonical partition function through
\begin{equation}
PV= k_B T \log \Theta(\mu,V,T)
\end{equation} 
As for the densities at coexistence, the electric field essentially shifts the curve for the vapor pressure towards the higher temperatures. For instance, in the absence of field, the vapor pressure reaches $50~bar$ around $550~K$. For $E=0.1~V/$\AA~, the vapor pressure reaches this value at a temperature of about $565~K$, i.e. $3~$\% above. Similarly, this value for the vapor pressure is achieved at temperatures larger by $7~$\% and by $12~$\%, for fields of $0.2~V/$\AA~ and $0.5~V/$\AA, respectively, than in the absence of field. If we now look at the effect of the field at fixed temperature, we see that the vapor pressure is decreased by $23~$\% if a field $E=0.1~V/$\AA~ is applied, by $47~$\% for $E=0.2~V/$\AA~ and by $65~$\% for $E=0.5~V/$\AA. This decrease in vapor pressure with the field intensity can be directly related to the decrease in the density of the vapor phase at coexistence observed for strong fields.

Fig.~\ref{Fig4} also shows the variation of the chemical potential at coexistence with temperature. This plot highlights the decrease in $\mu_{coex}$ with the applied field. For instance, at $T=575~K$ and using $\mu_{coex}$ in the absence of the field as the reference, we see that $\mu_{coex}$ is decreased by $3~$\% for $E=0.1~V/$\AA, by $7~$\% for $E=0.2~V/$\AA, and by $21~$\% for $E=0.5~V/$\AA. This shift towards the lower values of $\mu$ directly stems from the results obtained for the partition function (see Fig~\ref{Fig1}), which revealed that $\ln \Theta$ was increasingly shifted towards the lower end of the $\mu$ range as the applied field increased. The bottom panel in Fig.~\ref{Fig4} shows the dependence of the entropy of vaporization $\Delta S$ on the field. We find that the curve for $\Delta S$ against $T$ is shifted towards the top of the temperature range for strong fields. Using the results obtained in the absence of field at $T=575~K$ as the reference, we find an increase in $\Delta S$ of $22~$\% for $E=0.1~V/$\AA, of $35~$\% for $E=0.2~V/$\AA~and of $49~$\% for $E=0.5~V/$\AA. This can be attributed to the greater difference (in terms of density) between the two phases at coexistence for strong fields, as evidenced by the larger value for $(\rho_l -\rho_v)$ shown in the phase diagrams (see Fig.~\ref{Fig3}). However, another possible cause for this large $\Delta S$ may also be the structural changes induced by strong electric fields, a point that we aim to elucidate in the next paragraph.

\begin{figure}
\begin{center}
\includegraphics*[width=8cm]{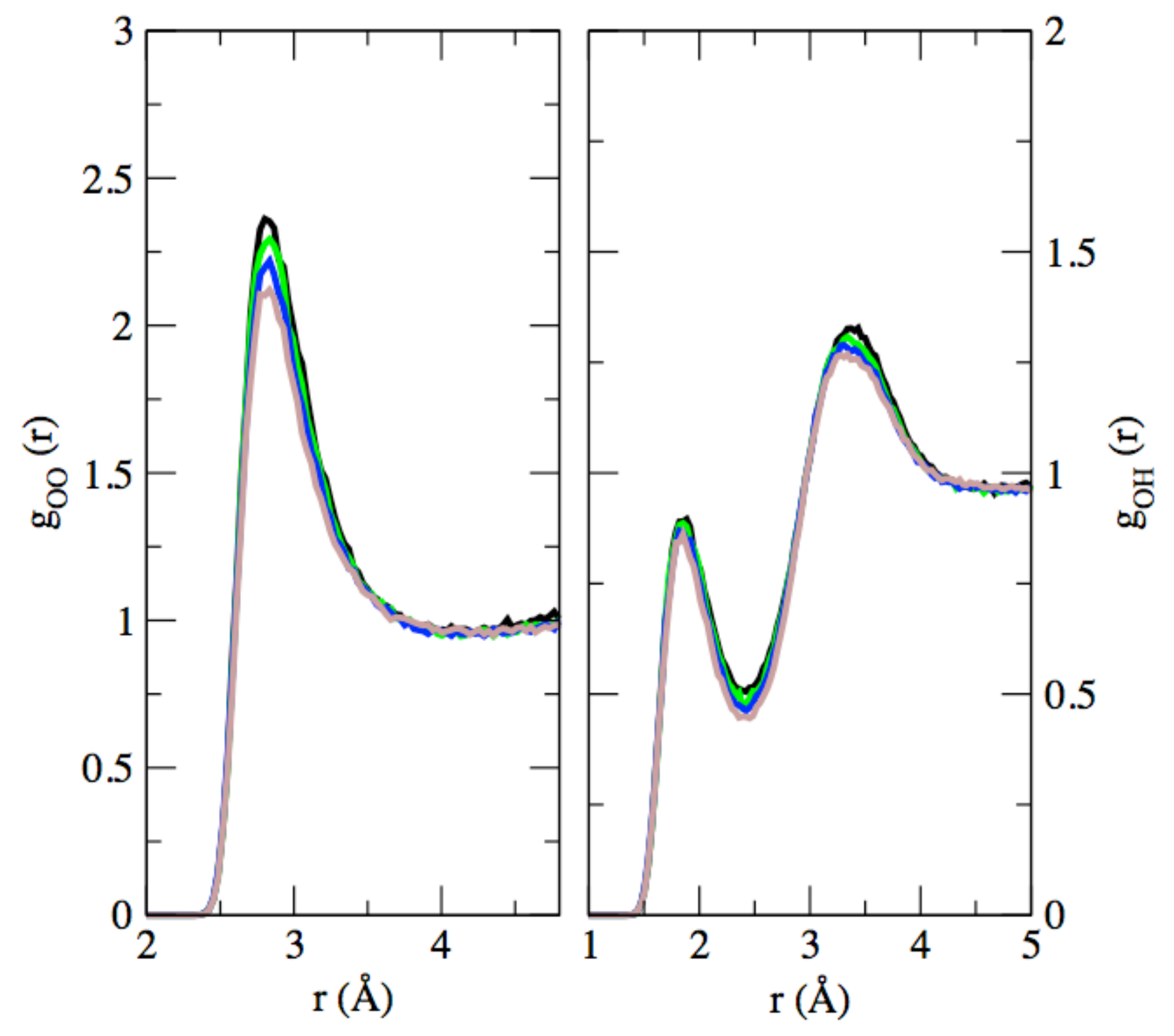}
\end{center}
\caption{Structural features of water at coexistence for $T=575$~K:  (Left) $g_{OO}(r)$ in the absence of field (black) and for field strengths of $E=0.1~V/$\AA~(green), $E=0.2~V/$\AA~(blue) and $E=0.5~V/$\AA~(brown), (Right) $g_{OO}(r)$ as a function of the applied field (same legend as for $g_{OO}(r)$).}
\label{Fig5}
\end{figure}

We focus here our analysis on the structure of the more organized phase, the saturated liquid, since the structure of vapor phases of water is known to be much less sensitive~\cite{SiepmannField}. We start by analyzing the radial distribution functions for the saturated liquid at $T=575~K$. We show in Fig.~\ref{Fig5} a comparison between the radial distribution functions $g_{OO}(r)$ obtained for different fields. The applied field only has a mild effect on this distribution function, which is consistent with prior results on low temperature liquids of water, which only revealed notable changes in radial distribution functions~\cite{gr1,gr2,gr3,gr4} for fields in excess of $0.5~V/$\AA. We are, however, able to notice a slight decrease in the height of the first peak in $g_{OO}(r)$ for high field strengths, which is reminiscent of the general effect of other external fields (e.g. shear) on these functions~\cite{theBook,EthanolShear}. We show in Fig.~\ref{Fig5} results for the $g_{OH}(r)$ distribution functions, which confirm the moderate effect of the field on the radial distribution functions~\cite{gr1,gr2,gr3,gr4} for fields up to $E=0.5~V/$\AA. This set of results would {\it a priori} indicate that the structure of the fluid has not been dramatically changed by the applied field. To ascertain this point, we refine our analysis by computing the average values for different order parameters for the saturated liquid.

We calculate the three following order parameters: $q_t$, $q_l$ and $N_H$. $q_t$ is tetrahedral order parameter which quantifies the amount of tetrahedrality among the four nearest neighbors of each water molecule~\cite{Chau,pdebene}. $q_t$ is calculated by averaging over all water molecules $i$ the local quantity defined as follows:
\begin{equation}
q_t(i)=1-\sum_{j=1} ^3 \sum_{k=j+1} ^4 {3 \over 8} \left( \phi_{jik} + {1 \over 3} \right)^2
\end{equation}
In this equation, $j$ and $k$ are two water molecules chosen among the four nearest neighbors of molecule $i$, and $\phi_{jik}$ is the angle between the vectors connecting $i$ and its two neighbors $j$ and $k$. $q_t$ is equal to $1$ if the environment around $i$ is perfectly tetrahedral. 

$q_l$ measures the alignment of water molecules in the direction of the electric field. It is calculated by averaging over all water molecules $i$ the following quantity
\begin{equation}
q_l(i)= {1 \over 2} \left( 3 \cos^2 \theta(i) -1 \right)
\end{equation}
where $\theta(i)$ is the angle between the electric field $\mathbf{E}$ and $\mathbf{m_i}$, the dipole moment of molecule $i$. $q_l$ reaches a value of $1$ when all dipoles are perfectly aligned with the field, a value of $0$ when there is no preferred orientation of the dipole and a value of $-0.5$ when the dipoles are perpendicular to the field.

$N_H$ is the number of hydrogen bonds within the saturated liquid. $N_H$ is obtained by applying the following 3 criteria~\cite{Hbond1,Hbond2,Hbond3,Hbond4}: (i) the $OO$ distance between two water molecules must be less than $3.5~$\AA~, (ii) the $O \cdots H$ distance between the $O$ of the first molecule and the $H$ of the second molecule involved in the hydrogen bond is less than $2.5~$\AA~ and (iii) the $O \cdots H-O$ angle along the hydrogen bond is less than $30^\circ$ 

\begin{figure}
\begin{center}
\includegraphics*[width=8cm]{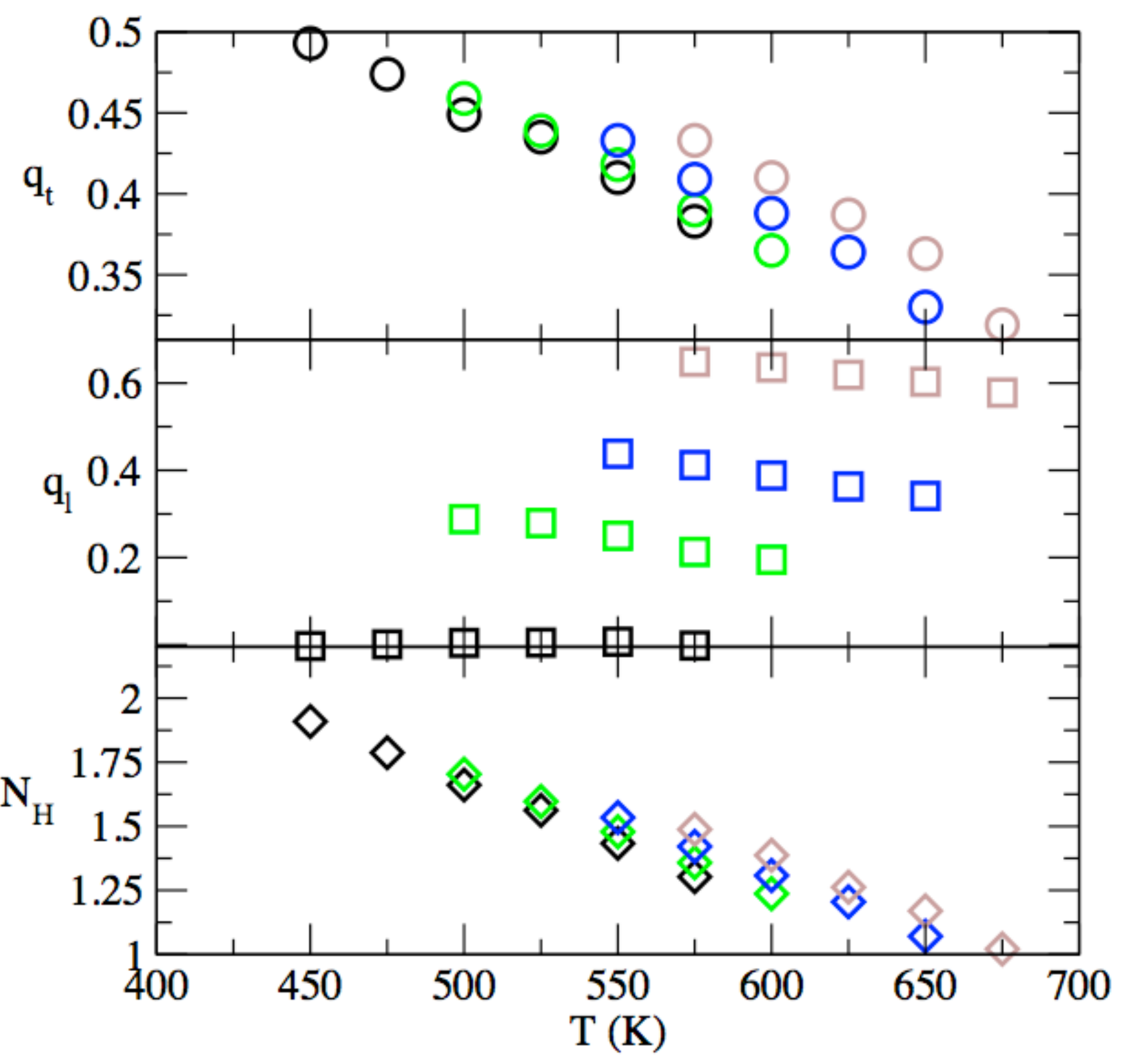}
\end{center}
\caption{Structural analysis for the saturated liquid along the coexistence line:  (From top to bottom) Tetrahedral order parameters $q_t$ (circles), alignment order parameter $q_l$ (squares) and number of hydrogen bonds $N_H$ (diamonds) in the absence of field (black) and for field strengths of $E=0.1~V/$\AA~(green), $E=0.2~V/$\AA~(blue) and $E=0.5~V/$\AA~(brown).}
\label{Fig6}
\end{figure}

We plot in Fig.~\ref{Fig6} the variations along the coexistence line of the 3 order parameters for the saturated liquid. Unlike the radial distribution functions that very weakly depended on the field, the order parameters all reveal that the amount of structural order in the liquid increases with the field. Fig.~\ref{Fig6} shows that an electric field of $0.1~V/$\AA~increases $q_t$ by $3~$\% and $N_H$ by $4~$\%. Applying a field of $0.2~V/$\AA~leads to an increase in $q_t$ of $7~$\% and of $N_H$ by $8~$\%. Similarly, a field of $0.5~V/$\AA~ results in an increase by $15~$\% for both $q_t$ and $N_H$. This increase in both $q_t$ and $N_H$ values concomitantly occurs with an increase in the alignment order parameter $q_l$ which goes from $0$ (no preferred orientation) in the absence of field to up to more than $0.6$ when the applied field if of $0.5~V/$\AA. This shows that the dramatic increase in the entropy of vaporization for strong fields (up to $49~$\% for $E=0.5~V/$\AA) therefore results from the cumulative effects of the greater density difference between the two coexisting phases and of the greater structural organization in the saturated liquid. 

\subsection{Liquid properties under an electric field}

\begin{figure}
\begin{center}
\includegraphics*[width=8cm]{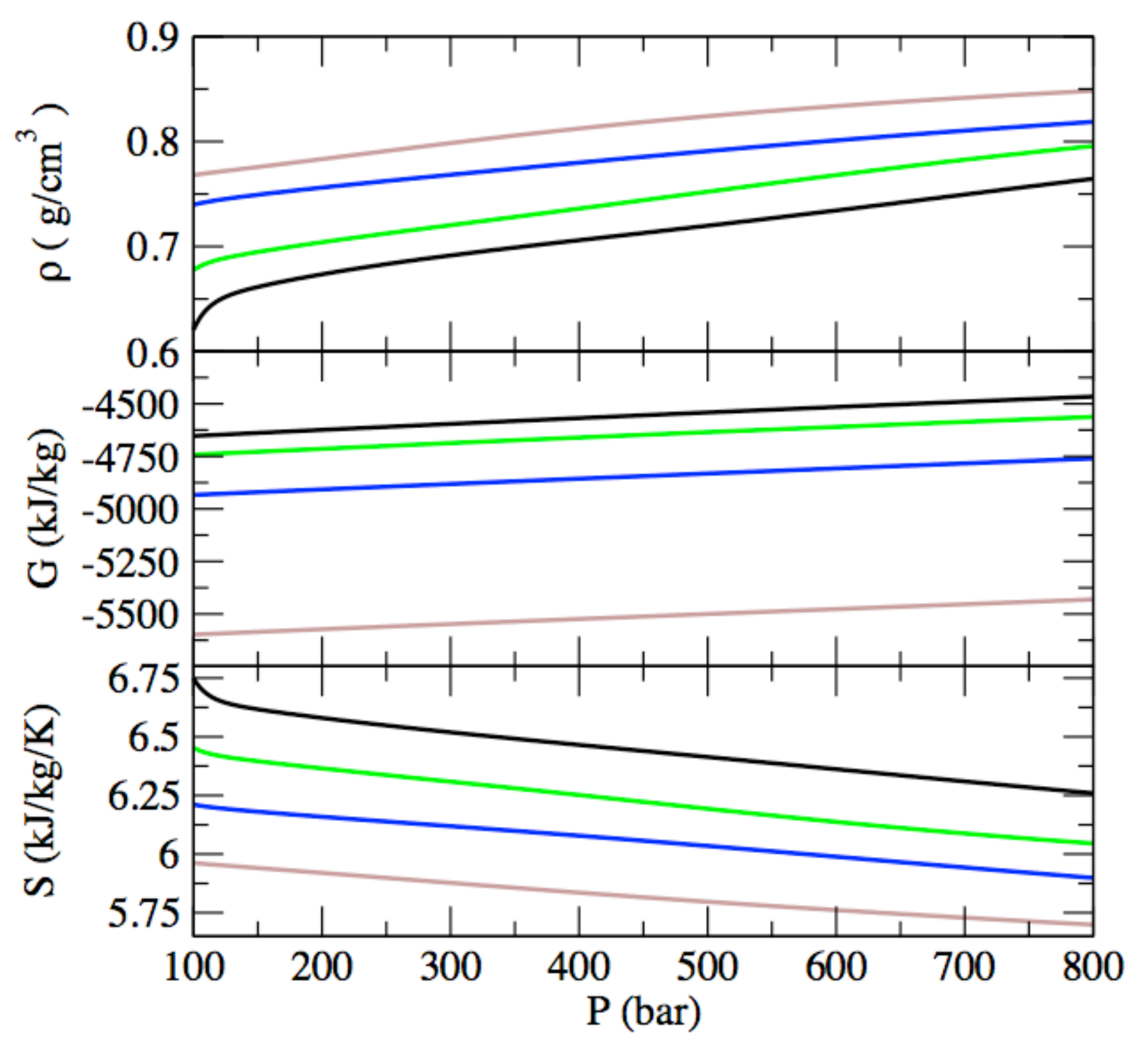}
\end{center}
\caption{Thermodynamic properties of compressed water at $575~K$. (Top) Density $\rho$ vs. pressure $P$ in the absence of field (black) and for field strengths of $E=0.1~V/$\AA~(green), $E=0.2~V/$\AA~(blue) and $E=0.5~V/$\AA~(brown), (Middle) Gibbs free energy $G$ vs. $P$, and (Bottom) Entropy $S$ vs. $P$.}
\label{Fig7}
\end{figure}

The results for the partition functions can also be used to shed light on the thermodynamics of compressed liquids. For this purpose, we vary the chemical potential $\mu$ and calculate the resulting number distribution $p(N)$ through Eq.~\ref{pN} and the thermodynamic properties from $p(N)$~\cite{PartI,PartIII}. We show in Fig.~\ref{Fig7} the results for the density $\rho$, the Gibbs free energy $G$ and the entropy $S$ of compressed liquids at $T=575~K$. In this plot, the variations of $\rho$, $G$ and $S$ with pressure are reported in the absence of field and for fields ranging from $0.1~V/$\AA~ to $0.5~V/$\AA. At fixed pressure, the dependence of the liquid density on the field is as follows: $\rho (0~V/$\AA$) < \rho (0.1~V/$\AA$) < \rho (0.2~V/$\AA$) < \rho (0.5~V/$\AA$)$. For instance at $P=500~bar$, the density for $E=0.1~V/$~\AA~ is $5~$\% greater than in the absence of field. Similarly, for $E=0.2~V/$\AA~, the density is greater by $10~$\% than in the absence of field, while for $E=0.5~V/$\AA~, the density is $15~$\% greater than in the absence of field. The increased density as a result of the applied field has two main consequences on liquid properties as evidenced by the behavior observed for $G$ and $S$ of the liquid. For $P=500~bar$ and using $G$ in the absence of field as the reference, $G$ is found to decrease by $2~$\% for $E=0.1~V/$\AA~, $7~$\% for $E=0.2~V/$\AA~ and $21~$\% for $E=0.5~V/$\AA. This decrease in $G$ can be attributed to the decrease in potential energy due to the larger number of attractive water-water interactions per unit volume and to the stronger interaction of water with the field, the latter becoming increasingly significant as the field gets stronger. We now focus on the variations of $S$ with the field. At $P=500~bar$, we find that $S$ decreases by $4~$\% when a field of $0.1~V/$\AA~ is applied, by $6~$\% when a field of $0.2~V/$\AA~ and by $10~$\% when a field of $0.5~V/$\AA. This decrease in $S$ can be attributed to the increase in the density of the fluid, but also to the greater organization within the compressed liquid arising from the applied field, as discussed below.

\begin{figure}
\begin{center}
\includegraphics*[width=8cm]{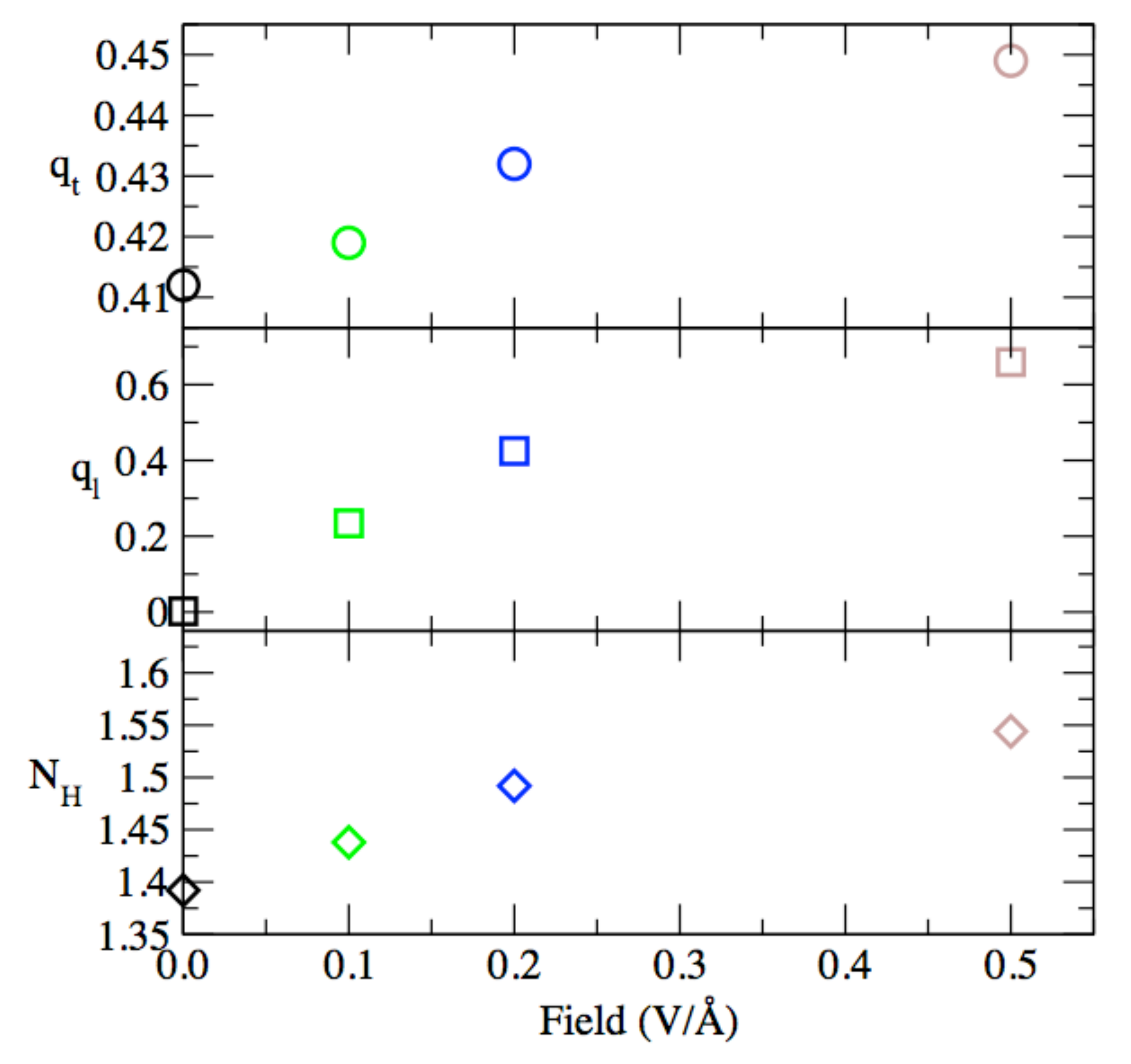}
\end{center}
\caption{Structural analysis for compressed liquids at $T=575K$ and $P=500$~bar:  (From top to bottom) Tetrahedral order parameters $q_t$ (circles), alignment order parameter $q_l$ (squares) and number of hydrogen bonds $N_H$ (diamonds) for the saturated liquid in the absence of field (black) and for field strengths of $E=0.1~V/$\AA~(green), $E=0.2~V/$\AA~(blue) and $E=0.5~V/$\AA~(brown).}
\label{Fig8}
\end{figure}

We plot in Fig.~\ref{Fig8} the results obtained for the three order parameters $q_t$, $q_l$ and $N_H$ at $P=500~bar$. All 3 order parameters point to a greater organization within the liquid as the field gets stronger.~$q_t$ is found to increase by $2~$\% for $E=0.1~V/$\AA~, by $5~$\% for $E=0.2~V/$\AA~ and by $9~$\% for $E=0.5~V/$\AA. The degree of alignment also steadily increases with the field and shows that $q_l$ goes from $0$ in the absence of field to up to $0.66$ for $E=0.5~V/$\AA. The number of hydrogen bonds within the compressed liquid is also found to increase with the field. With respect to the liquid at $P=500~bar$ in the absence of field, we find that $N_H$ is greater by $3~$\% for $E=0.1~V/$\AA~, by  $7~$\% for $E=0.2~V/$\AA~ and by $11~$\% for $E=0.5~V/$\AA. These results, together with the increase in density with the field, account for the observed decrease in entropy for strong fields.

\subsection{Supercritical water under an electric field}

\begin{figure}
\begin{center}
\includegraphics*[width=8cm]{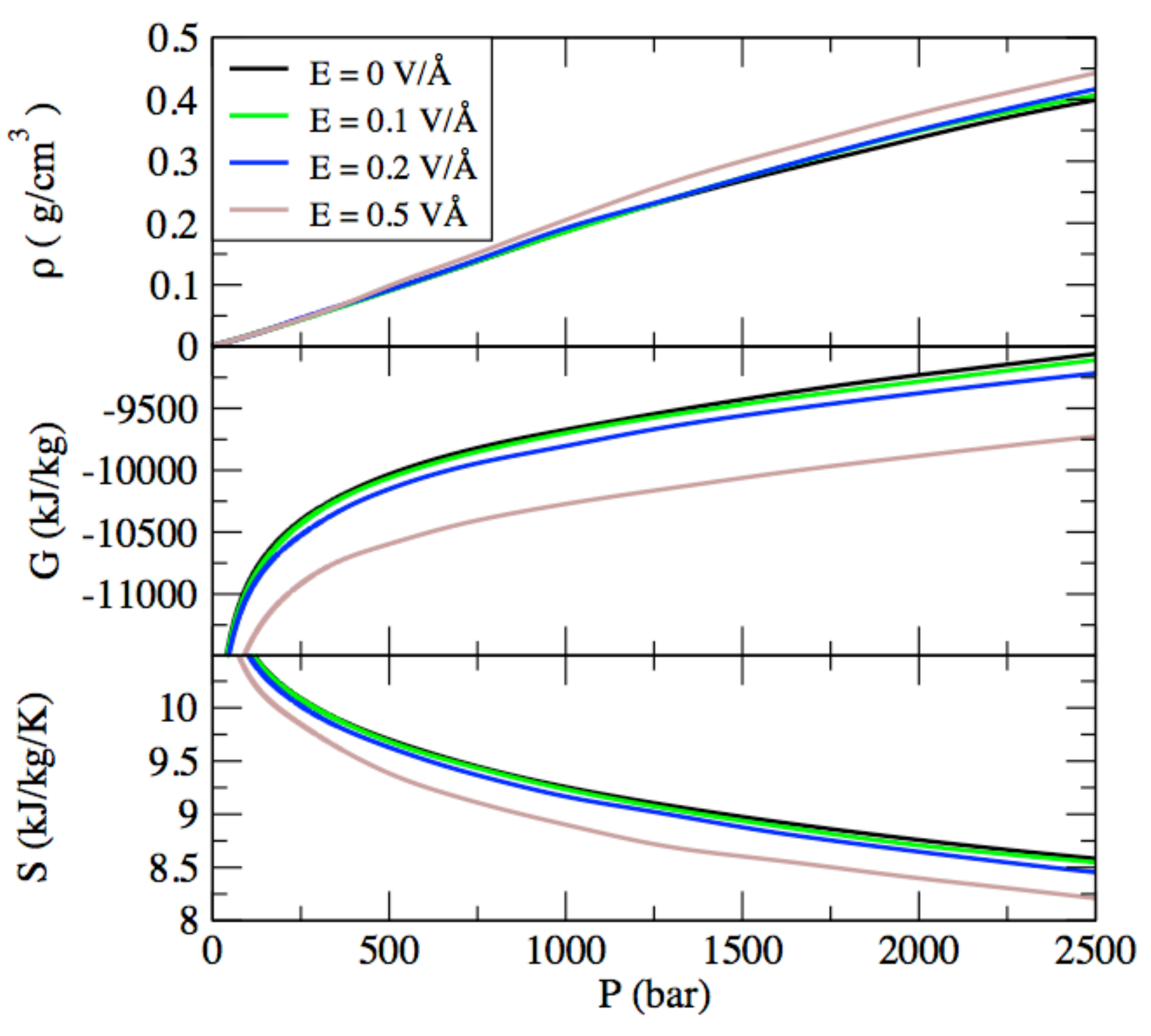}
\end{center}
\caption{Thermodynamic properties of supercritical water at $1250~K$. (Top) Density $\rho$ vs. pressure $P$ in the absence of field and for different field strengths, (Middle) Gibbs free energy $G$ vs. $P$, and (Bottom) Entropy $S$ vs. $P$.}
\label{Fig9}
\end{figure}

\begin{figure}
\begin{center}
\includegraphics*[width=8cm]{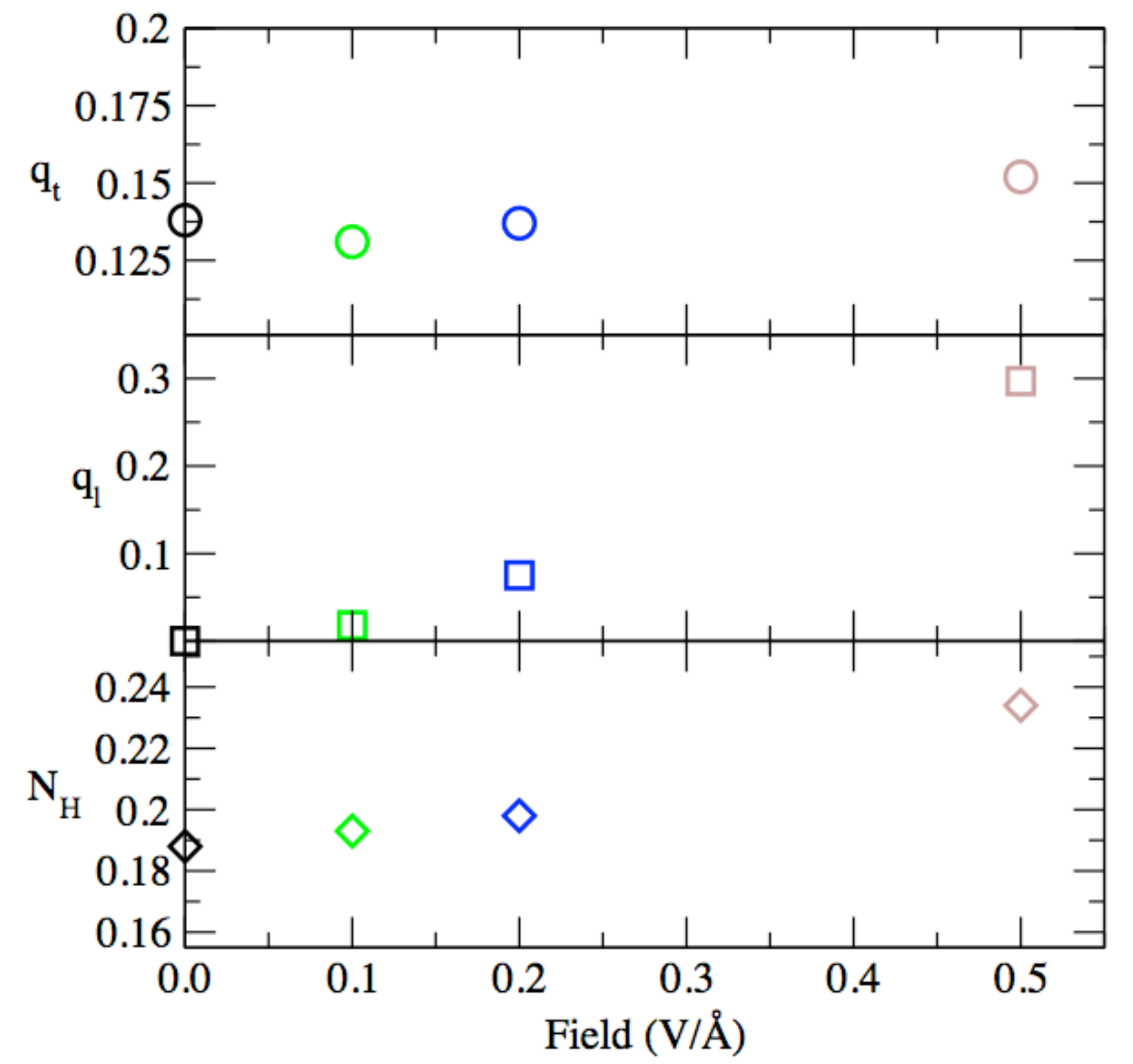}
\end{center}
\caption{Structural analysis for supercritical fluids at $T=1250K$ and $P=1000$~bar:  (From top to bottom) Tetrahedral order parameters $q_t$ (circles), alignment order parameter $q_l$ (squares) and number of hydrogen bonds $N_H$ (diamonds) for the saturated liquid in the absence of field (black) and for field strengths of $E=0.1~V/$\AA~(green), $E=0.2~V/$\AA~(blue) and $E=0.5~V/$\AA~(brown).}
\label{Fig10}
\end{figure}

We now examine the impact of the field on the thermodynamic properties of supercritical water. Fig.~\ref{Fig9} shows plots of $\rho$, $G$ and $S$ against pressure. The impact of the field on the density of supercritical fluids is more limited than for liquids. For instance, for $P=1000~bar$, the increase in density only starts to become noticeable for $E=0.2~V/$\AA, which results in a density increase of $3~$\% with respect to the density in the absence of field. Increasing further the field to $0.5~V/$\AA~yields a density $10~$\% greater than in the absence of field. The impact on $G$ and $S$ follows a similar trend. For $G$, increasing the field from $0~V/$\AA~to $0.2~V/$\AA~decreases the $G$ by about $1.5~$\% and an increase in the field from $0~V/$\AA~to $0.5~V/$\AA~decreases $G$ by $6~$\%. As for the liquid, we attribute this decrease in $G$ to the greater number of attractive water-water interactions and to the greater interaction energy with the field. For $S$, decreases in entropy with respect to the value in the absence of field are of $1~$\% and $4~$\% for fields of $0.2~V/$\AA~and $0.5~V/$\AA, respectively. This stems from the increase in density with the field and from the increase in structural order within the fluid (see Fig.~\ref{Fig10}). Fig.~\ref{Fig10} shows that, $E=0.5~V/$\AA, for  $q_t$ is e.g. greater by $10~$\% than in he absence of field, that $q_l$ reaches up to $0.3$ and that $N_H$ increases by $24~$\%.

We continue our characterization of the thermodynamics of supercritical water with the determination of the ideality contours for water. These contours~\cite{nedostup2013asymptotic,apfelbaum2013regarding} have recently emerged as a means to bridge the gap in our understanding of the supercritical region of the phase diagram~\cite{brazhkin2013liquid,brazhkin2011van,Morita,brazhkin2012supercritical}, since they provide a way to map these regions and to establish a correspondence between the supercritical states of different fluids~\cite{Abigail}. Here we focus on two of these contours. We start by determining the locus for the Zeno line, defined as the contour along which supercritical water behaves as an ideal gas from the standpoint of the ideal gas law. This locus is obtained by varying numerically $\mu$ such that the following condition is satisfied:
\begin{equation}
P\bar{V}/RT= \log \Theta(\mu,V,T) /\bar{N}=1
\label{Zeno}
\end{equation} 
where $\bar{V}$ is the reciprocal density and $\bar{N}=\sum N p(N)$ is the average number of particles in the system. Previous work in the field has focused on establishing the shape of this contour for the Van der Waals equation~\cite{nedostup2013asymptotic}, model systems~\cite{apfelbaum2009confirmation}, Argon~\cite{apfelbaum2013regarding}, metals~\cite{apfelbaum2009correspondence,apfelbaum2015similarity,Leanna,Apfelbaum2016} and a few molecular fluids including water~\cite{kutney2000zeno,Abigail}. Remarkably, it has been shown that the Zeno line is a straight line that extends over several hundred degrees. However, the effect of an electric field on the shape of the Zeno line has not been studied so far, and it remains to be seen how the field impacts the shape of this contour. 

The second contour we study in this work is the $H$ line, defined as the curve of ideal enthalpy~\cite{nedostup2013asymptotic,apfelbaum2013regarding}. We obtain the locus for this contour by varying numerically $\mu$ such that the condition written below is obeyed.
\begin{equation}
\bar H = \bar U + RT \log \Theta(\mu,V,T) /\bar{N} = 4 RT
\label{Hline}
\end{equation}
Similarly to the Zeno line, the locus for this contour has been shown to correspond to a straight line for the Van der Waals equation~\cite{nedostup2013asymptotic}, Argon~\cite{apfelbaum2013regarding} and several molecular fluids~\cite{Abigail}. It has, however, been studied much less extensively than the Zeno line and, to our knowledge, the impact of an electric field on the $H$ line has yet to be investigated.

\begin{figure}
\begin{center}
\includegraphics*[width=8cm]{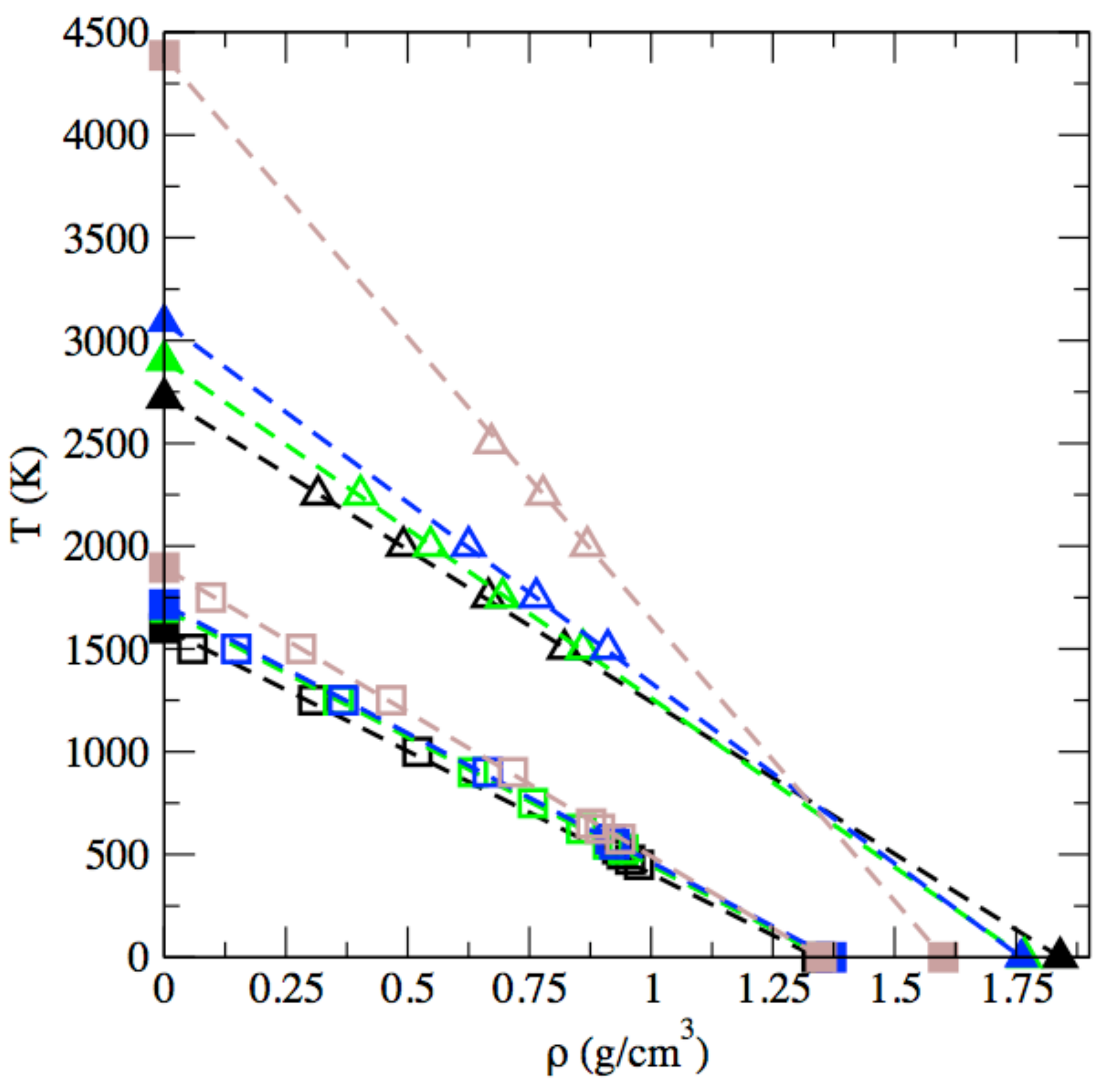}
\end{center}
\caption{Zeno line and $H$ line for water in the absence of field (black) and for field strengths of $E=0.1~V/$\AA~(green), $E=0.2~V/$\AA~(blue) and $E=0.5~V/$\AA~(brown).}
\label{Fig11}
\end{figure}

We report in Fig.~\ref{Fig11} the sets of $(T, \rho)$ satisfying the conditions given in Eqs.~\ref{Zeno} and~\ref{Hline} for water in the absence of field and for fields ranging from $0.1~V/$\AA~to $0.5~V/$\AA. We then fit the EWL results to linear laws (also shown in Fig.~\ref{Fig11}). Both the Zeno and $H$ line remain remarkably straight lines regardless of the field strength. To establish further this point, we evaluate the departure of the Zeno lines from straight lines by calculating the Average Absolute Relative Deviation ($AARD$) error ($AARD(\%)= {{{1}\over{N}}\sum_i {\left|{T_{Fit}-T_{sim}} \over{T_{sim}} \right|}}$). For all fields, the AARD remains small (between $1$\% and $3$\%) and of the same order as in the absence of field ($3.7$\%). This means that the shape of the Zeno line is not altered by the electric field and remains the same as in the absence of the field, thus opening the door for the application of the similarity laws based on the Zeno line~\cite{apfelbaum2009correspondence} to systems under an electric field. As shown in Fig.~\ref{Fig11}, only high temperature results are available for the $H$ line, since the low temperature/high density domain for the $H$ line lie within the domain of stability of the solid. These linear fits allow us to determine the Boyle and $H$ parameters, which are key input parameters in the similarity laws of Apfelbaum and Vorob'ev~\cite{apfelbaum2008new}. The Boyle parameters are obtained as the intercept of the Zeno line with the temperature axis (this gives the Boyle temperature $T_B$) and with the density axis (this provides the Boyle density $\rho_B$). The results for the Boyle parameters are summarized in Table~\ref{param}. While the Boyle density remains essentially constant throughout the range of fields considered in this work, the Boyle temperature steadily increases with the field strength. $T_B$ is greater by $6~$\% for $E=0.1~V/$\AA~ than in the absence of field, by $7.5~$\% for $E=0.2~V/$\AA~and by $18.5~$\% for $E=0.5~V/$\AA. Similarly, the intercepts of the $H$ line with the temperature and density axes provide the two parameters $T_H$ and $\rho_H$ (given in Table~\ref{param}). In line with the results obtained for $T_c$ and $T_B$, $T_H$ is shown to increase with the field, with a $7~$\% increase for $E=0.1~V/$\AA, a $13~$\% increase for $E=0.2~V/$\AA~and  a $61~$\% increase for $E=0.5~V/$\AA. As observed for $\rho_c$, $\rho_H$ is found to decrease with the field with a $4~$\% decrease for $E=0.1~V/$\AA~and a $13$\% increase for $E=0.5~V/$\AA~(both with respect to the value for $T_H$ in the absence of field).

\begin{table}[hbpt]
\caption{Boyle, $H$ and critical parameters for $H_2O$ under an electric field ($E$ are given in $V/$\AA, $T$ in $K$ and $\rho$ in $g/cm^3$). Uncertainties are within $1$\% for $T_c$ and $2$\% for $\rho_c$, within $1$\% for $T_B$ and $2$\% for $\rho_B$,  and within $0.8$\% for $T_H$ and $2$\% for $\rho_H$.}
\begin{tabular}{|c|c|c|c|c|c|c|c|}
\hline
\hline
$E$ &  $T_B$ &  $\rho_B$ & $T_H$ & $\rho_H$ & $T_c$ & $\rho_c$\\
\hline
\hline
0 & 1599 & 1.343 & 2722 & 1.844 & 641 & 0.310\\
0.05 & 1644 & 1.363 & 2801 & 1.766 & 643 & 0.304\\ 
0.1 & 1694 & 1.358 & 2904 & 1.766 & 660 & 0.303\\
0.2 & 1716 & 1.374 & 3090 & 1.768 & 683 & 0.298\\
0.5 & 1895 & 1.344 & 4387 & 1.597 & 706 & 0.291\\
\hline
\hline
\end{tabular}
\label{param}
\end{table}

\begin{figure}
\begin{center}
\includegraphics*[width=8cm]{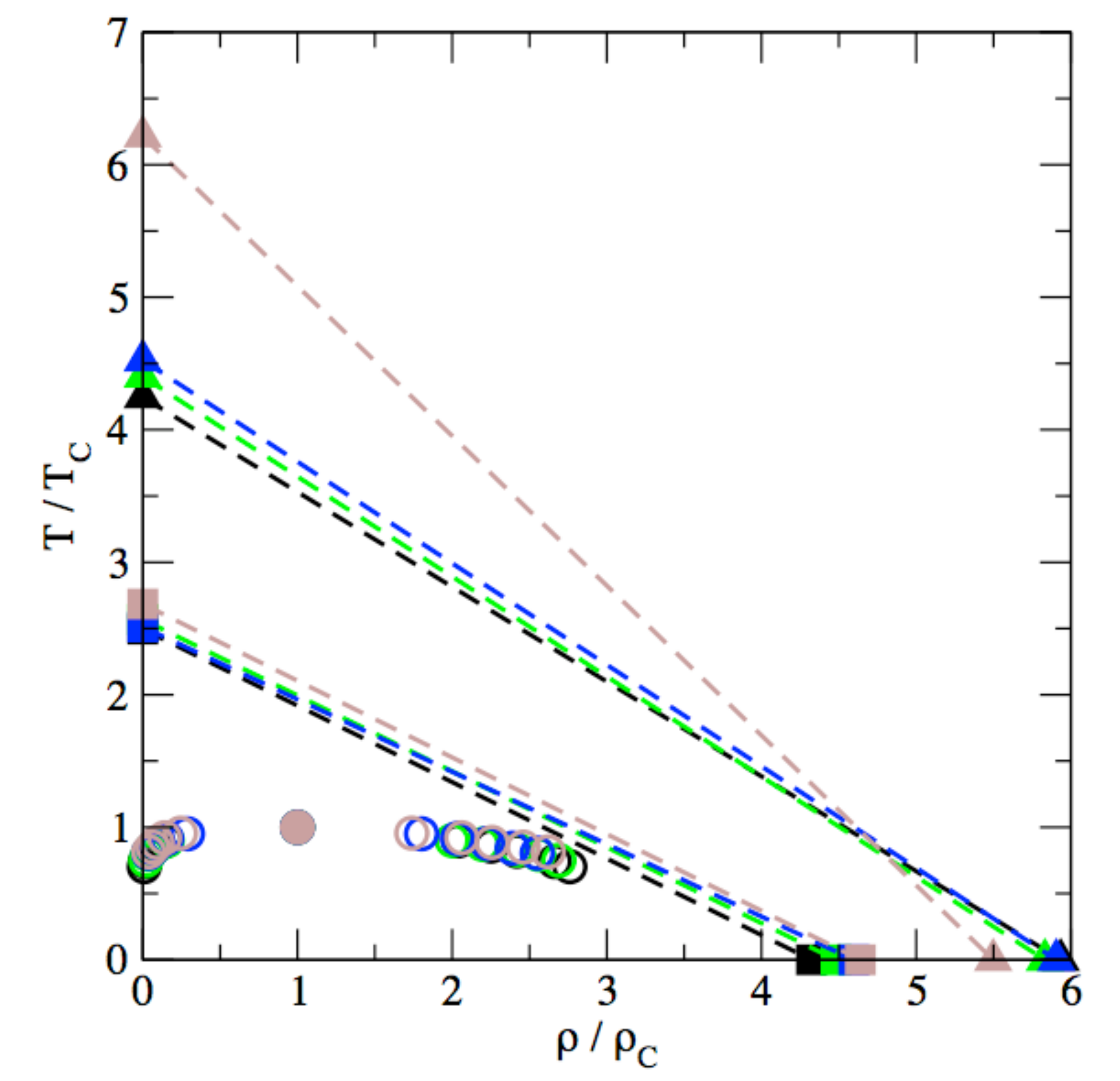}
\end{center}
\caption{Scaled phase diagram and ideality contours for water in the absence of field (black) and for field strengths of $E=0.1~V/$\AA~(green), $E=0.2~V/$\AA~(blue) and $E=0.5~V/$\AA~(brown). Results for each value of the field are scaled by their respective critical temperature and critical density.}
\label{Fig12}
\end{figure}

To provide a comparison between the binodal curve and ideality contours obtained with and without the field, we rescale the phase diagram of water by the critical temperature and critical density found for each value of the field. Fig.~\ref{Fig12}~shows that the binodals obtained for all fields can all be superimposed on top of one another. This shows that the behavior of subcritical water remains qualitatively the same for very strong fields (up to $0.5~V/$\AA). Similarly, the loci obtained for the Zeno line in the scaled temperature-density plane are in good agreement for all fields, with maximum deviations of less than $7~$\% for the scaled $T_B$ and $\rho_B$ for all fields. Similar conclusions only apply for the $H$ line for fields of up to $0.2~V/$\AA. This can best be seen on the scaled $T_H$ for $E=0.5~V/$\AA. For this field, the scaled $T_H$ is $46~$\% greater than in the absence of field and exhibits a markedly larger value than for all other conditions. This shows that while a correspondence between the results obtained for different fields can be made on the basis of the scaled phase diagrams for fields of up to $0.2~V/$\AA, the behavior of water subjected to fields of $0.5~V/$\AA~and beyond is qualitatively different. This discrepancy observed in the $H$ line for the strongest field is likely due to the predominant contribution of the potential energy due to the interaction of water molecules with the applied field, and the strong alignment of water molecules with the field, along the $H$ line in these low density-supercritical states.

\section{Conclusion}
In this work, we extend the Expanded Wang-Landau simulation method to determine the impact of an electric field on the phase diagram of water, on the thermodynamic properties of the vapor-liquid transition, of compressed liquids and of supercritical phases of water, as well as on the loci for the ideality contours of water. Through staged insertions/deletions of water molecules, the EWL method allows us to calculate the grand-canonical partition function of water under an electric field and to determine its properties using the statistical mechanics formalism. Our results show that the impact of the electric field on the partition function becomes significant for fields greater than $0.05~V/$\AA~ and that it steadily increases with the field. This, in turn, has a number of important consequences on the phase diagram with a shift of the binodal towards the lower densities/higher temperatures as the field gets stronger. This also leads to a decrease in the chemical potential at coexistence (by up to $21~$\%) and to a significant increase in the entropy of vaporization (by up to $49~$\%). This result is attributed to the greater difference between the two phases at coexistence, both in terms of density and structural order, as shown by the analysis of the dependence on the field of the tetrahedral order parameter, of the extent of the alignment of water molecules with the field and of the number of hydrogen bonds. These conclusion extend to the thermodynamics of compressed liquids and of supercritical phases of water. Finally, a correspondence between the results obtained for different field strength is carried out through the analysis of the ideality contours. The results show that the Zeno and $H$ lines remain straight over the entire range of electric fields studied in this work. This correspondence, however, starts to break down for a field of $0.5~V/$\AA, which lead to markedly different $H$ parameters as a result of the predominance of the field-water interaction in these low density-supercritical states. Further work using more sophisticated models for water, including e.g. polarizable models, quantum effects~\cite{ceriotti2016nuclear} and allowing for water dissociation~\cite{saitta2012ab,geissler2001autoionization,chau2001electrical,chau2011chemical}, will allow for a refinement of these findings.

{\bf Acknowledgements}

Partial funding for this research was provided by NSF through CAREER award DMR-1052808.


\begin{thebibliography}{96}
\expandafter\ifx\csname natexlab\endcsname\relax\def\natexlab#1{#1}\fi
\expandafter\ifx\csname bibnamefont\endcsname\relax
  \def\bibnamefont#1{#1}\fi
\expandafter\ifx\csname bibfnamefont\endcsname\relax
  \def\bibfnamefont#1{#1}\fi
\expandafter\ifx\csname citenamefont\endcsname\relax
  \def\citenamefont#1{#1}\fi
\expandafter\ifx\csname url\endcsname\relax
  \def\url#1{\texttt{#1}}\fi
\expandafter\ifx\csname urlprefix\endcsname\relax\def\urlprefix{URL }\fi
\providecommand{\bibinfo}[2]{#2}
\providecommand{\eprint}[2][]{\url{#2}}

\bibitem[{\citenamefont{Siria et~al.}(2013)\citenamefont{Siria, Poncharal,
  Biance, Fulcrand, Blase, Purcell, and Bocquet}}]{nano1}
\bibinfo{author}{\bibfnamefont{A.}~\bibnamefont{Siria}},
  \bibinfo{author}{\bibfnamefont{P.}~\bibnamefont{Poncharal}},
  \bibinfo{author}{\bibfnamefont{A.-L.} \bibnamefont{Biance}},
  \bibinfo{author}{\bibfnamefont{R.}~\bibnamefont{Fulcrand}},
  \bibinfo{author}{\bibfnamefont{X.}~\bibnamefont{Blase}},
  \bibinfo{author}{\bibfnamefont{S.~T.} \bibnamefont{Purcell}},
  \bibnamefont{and} \bibinfo{author}{\bibfnamefont{L.}~\bibnamefont{Bocquet}},
  \bibinfo{journal}{Nature} \textbf{\bibinfo{volume}{494}},
  \bibinfo{pages}{455} (\bibinfo{year}{2013}).

\bibitem[{\citenamefont{Janssen and Pennathur}(2015)}]{nano2}
\bibinfo{author}{\bibfnamefont{K.~G.} \bibnamefont{Janssen}} \bibnamefont{and}
  \bibinfo{author}{\bibfnamefont{S.}~\bibnamefont{Pennathur}},
  \bibinfo{journal}{Lab on a Chip} \textbf{\bibinfo{volume}{15}},
  \bibinfo{pages}{3980} (\bibinfo{year}{2015}).

\bibitem[{\citenamefont{Rinne et~al.}(2012)\citenamefont{Rinne, Gekle,
  Bonthuis, and Netz}}]{nano3}
\bibinfo{author}{\bibfnamefont{K.~F.} \bibnamefont{Rinne}},
  \bibinfo{author}{\bibfnamefont{S.}~\bibnamefont{Gekle}},
  \bibinfo{author}{\bibfnamefont{D.~J.} \bibnamefont{Bonthuis}},
  \bibnamefont{and} \bibinfo{author}{\bibfnamefont{R.~R.} \bibnamefont{Netz}},
  \bibinfo{journal}{Nano Lett.} \textbf{\bibinfo{volume}{12}},
  \bibinfo{pages}{1780} (\bibinfo{year}{2012}).

\bibitem[{\citenamefont{De~Luca et~al.}(2013)\citenamefont{De~Luca, Todd,
  Hansen, and Daivis}}]{nano4}
\bibinfo{author}{\bibfnamefont{S.}~\bibnamefont{De~Luca}},
  \bibinfo{author}{\bibfnamefont{B.}~\bibnamefont{Todd}},
  \bibinfo{author}{\bibfnamefont{J.}~\bibnamefont{Hansen}}, \bibnamefont{and}
  \bibinfo{author}{\bibfnamefont{P.~J.} \bibnamefont{Daivis}},
  \bibinfo{journal}{J. Chem. Phys.} \textbf{\bibinfo{volume}{138}},
  \bibinfo{pages}{154712} (\bibinfo{year}{2013}).

\bibitem[{\citenamefont{Corovic et~al.}(2013)\citenamefont{Corovic, Lackovic,
  Sustaric, Sustar, Rodic, and Mikalvcic}}]{bio1}
\bibinfo{author}{\bibfnamefont{S.}~\bibnamefont{Corovic}},
  \bibinfo{author}{\bibfnamefont{I.}~\bibnamefont{Lackovic}},
  \bibinfo{author}{\bibfnamefont{P.}~\bibnamefont{Sustaric}},
  \bibinfo{author}{\bibfnamefont{T.}~\bibnamefont{Sustar}},
  \bibinfo{author}{\bibfnamefont{T.}~\bibnamefont{Rodic}}, \bibnamefont{and}
  \bibinfo{author}{\bibfnamefont{D.}~\bibnamefont{Mikalvcic}},
  \bibinfo{journal}{Biomedical Engineering Online}
  \textbf{\bibinfo{volume}{12}}, \bibinfo{pages}{16} (\bibinfo{year}{2013}).

\bibitem[{\citenamefont{Wikstrom et~al.}(2003)\citenamefont{Wikstrom,
  Verkhovsky, and Hummer}}]{bio2}
\bibinfo{author}{\bibfnamefont{M.}~\bibnamefont{Wikstrom}},
  \bibinfo{author}{\bibfnamefont{M.~I.} \bibnamefont{Verkhovsky}},
  \bibnamefont{and} \bibinfo{author}{\bibfnamefont{G.}~\bibnamefont{Hummer}},
  \bibinfo{journal}{Biochimica et Biophysica Acta}
  \textbf{\bibinfo{volume}{1604}}, \bibinfo{pages}{61} (\bibinfo{year}{2003}).

\bibitem[{\citenamefont{Tsouris et~al.}(2001)\citenamefont{Tsouris, Blakenship,
  Dong, and DePaoli}}]{dist1}
\bibinfo{author}{\bibfnamefont{C.}~\bibnamefont{Tsouris}},
  \bibinfo{author}{\bibfnamefont{K.~D.} \bibnamefont{Blakenship}},
  \bibinfo{author}{\bibfnamefont{J.}~\bibnamefont{Dong}}, \bibnamefont{and}
  \bibinfo{author}{\bibfnamefont{D.~W.} \bibnamefont{DePaoli}},
  \bibinfo{journal}{Ind. Eng. Chem. Res.} \textbf{\bibinfo{volume}{40}},
  \bibinfo{pages}{3843} (\bibinfo{year}{2001}).

\bibitem[{\citenamefont{amd S.~Pan et~al.}(2001)\citenamefont{amd S.~Pan, Liu,
  and Dai}}]{dist2}
\bibinfo{author}{\bibfnamefont{L.~L.} \bibnamefont{amd S.~Pan}},
  \bibinfo{author}{\bibfnamefont{J.~G.} \bibnamefont{Liu}}, \bibnamefont{and}
  \bibinfo{author}{\bibfnamefont{Y.~Y.} \bibnamefont{Dai}},
  \bibinfo{journal}{Sep. Sci. Technol.} \textbf{\bibinfo{volume}{36}},
  \bibinfo{pages}{2799} (\bibinfo{year}{2001}).

\bibitem[{\citenamefont{Kirkby et~al.}(2016)\citenamefont{Kirkby, Duplissy,
  Sengupta, Frege, Gordon, Williamson, Heinritzi, Simon, Yan, Almeida
  et~al.}}]{atm1}
\bibinfo{author}{\bibfnamefont{J.}~\bibnamefont{Kirkby}},
  \bibinfo{author}{\bibfnamefont{J.}~\bibnamefont{Duplissy}},
  \bibinfo{author}{\bibfnamefont{K.}~\bibnamefont{Sengupta}},
  \bibinfo{author}{\bibfnamefont{C.}~\bibnamefont{Frege}},
  \bibinfo{author}{\bibfnamefont{H.}~\bibnamefont{Gordon}},
  \bibinfo{author}{\bibfnamefont{C.}~\bibnamefont{Williamson}},
  \bibinfo{author}{\bibfnamefont{M.}~\bibnamefont{Heinritzi}},
  \bibinfo{author}{\bibfnamefont{M.}~\bibnamefont{Simon}},
  \bibinfo{author}{\bibfnamefont{C.}~\bibnamefont{Yan}},
  \bibinfo{author}{\bibfnamefont{J.}~\bibnamefont{Almeida}},
  \bibnamefont{et~al.}, \bibinfo{journal}{Nature}
  \textbf{\bibinfo{volume}{533}}, \bibinfo{pages}{521} (\bibinfo{year}{2016}).

\bibitem[{\citenamefont{Fisenko et~al.}(2005)\citenamefont{Fisenko, Kane, and
  El-Shall}}]{atm2}
\bibinfo{author}{\bibfnamefont{S.~P.} \bibnamefont{Fisenko}},
  \bibinfo{author}{\bibfnamefont{D.~B.} \bibnamefont{Kane}}, \bibnamefont{and}
  \bibinfo{author}{\bibfnamefont{M.~S.} \bibnamefont{El-Shall}},
  \bibinfo{journal}{J. Chem. Phys.} \textbf{\bibinfo{volume}{123}},
  \bibinfo{pages}{104704} (\bibinfo{year}{2005}).

\bibitem[{\citenamefont{Kathmann et~al.}(2005)\citenamefont{Kathmann, Schenter,
  and Garrett}}]{atm3}
\bibinfo{author}{\bibfnamefont{S.~M.} \bibnamefont{Kathmann}},
  \bibinfo{author}{\bibfnamefont{G.~K.} \bibnamefont{Schenter}},
  \bibnamefont{and} \bibinfo{author}{\bibfnamefont{B.~C.}
  \bibnamefont{Garrett}}, \bibinfo{journal}{Phys. Rev. Lett.}
  \textbf{\bibinfo{volume}{94}}, \bibinfo{pages}{116104}
  (\bibinfo{year}{2005}).

\bibitem[{\citenamefont{Svishchev and Kusalik}(1999)}]{pgku}
\bibinfo{author}{\bibfnamefont{I.}~\bibnamefont{Svishchev}} \bibnamefont{and}
  \bibinfo{author}{\bibfnamefont{P.}~\bibnamefont{Kusalik}},
  \bibinfo{journal}{J. Am. Chem. Soc.} \textbf{\bibinfo{volume}{118}},
  \bibinfo{pages}{649} (\bibinfo{year}{1999}).

\bibitem[{\citenamefont{Zhou et~al.}(2013)\citenamefont{Zhou, Liu, Zhu, Lin,
  Pan, Jing, and Wang}}]{pol1}
\bibinfo{author}{\bibfnamefont{Y.~S.} \bibnamefont{Zhou}},
  \bibinfo{author}{\bibfnamefont{Y.}~\bibnamefont{Liu}},
  \bibinfo{author}{\bibfnamefont{G.}~\bibnamefont{Zhu}},
  \bibinfo{author}{\bibfnamefont{Z.-H.} \bibnamefont{Lin}},
  \bibinfo{author}{\bibfnamefont{C.}~\bibnamefont{Pan}},
  \bibinfo{author}{\bibfnamefont{Q.}~\bibnamefont{Jing}}, \bibnamefont{and}
  \bibinfo{author}{\bibfnamefont{Z.~L.} \bibnamefont{Wang}},
  \bibinfo{journal}{Nano Lett.} \textbf{\bibinfo{volume}{13}},
  \bibinfo{pages}{2771} (\bibinfo{year}{2013}).

\bibitem[{\citenamefont{Zhou et~al.}(2014)\citenamefont{Zhou, Wang, Yang, Zhu,
  Niu, Lin, Liu, and Wang}}]{pol2}
\bibinfo{author}{\bibfnamefont{Y.~S.} \bibnamefont{Zhou}},
  \bibinfo{author}{\bibfnamefont{S.}~\bibnamefont{Wang}},
  \bibinfo{author}{\bibfnamefont{Y.}~\bibnamefont{Yang}},
  \bibinfo{author}{\bibfnamefont{G.}~\bibnamefont{Zhu}},
  \bibinfo{author}{\bibfnamefont{S.}~\bibnamefont{Niu}},
  \bibinfo{author}{\bibfnamefont{Z.-H.} \bibnamefont{Lin}},
  \bibinfo{author}{\bibfnamefont{Y.}~\bibnamefont{Liu}}, \bibnamefont{and}
  \bibinfo{author}{\bibfnamefont{Z.~L.} \bibnamefont{Wang}},
  \bibinfo{journal}{Nano Lett.} \textbf{\bibinfo{volume}{14}},
  \bibinfo{pages}{1567} (\bibinfo{year}{2014}).

\bibitem[{\citenamefont{Xi et~al.}(2012)\citenamefont{Xi, Zhao, Tong, and
  Cao}}]{pol3}
\bibinfo{author}{\bibfnamefont{X.}~\bibnamefont{Xi}},
  \bibinfo{author}{\bibfnamefont{D.}~\bibnamefont{Zhao}},
  \bibinfo{author}{\bibfnamefont{F.}~\bibnamefont{Tong}}, \bibnamefont{and}
  \bibinfo{author}{\bibfnamefont{T.}~\bibnamefont{Cao}}, \bibinfo{journal}{Soft
  Matter} \textbf{\bibinfo{volume}{8}}, \bibinfo{pages}{298}
  (\bibinfo{year}{2012}).

\bibitem[{\citenamefont{Stepanow and Thurn-Albrecht}(2009)}]{down1}
\bibinfo{author}{\bibfnamefont{S.}~\bibnamefont{Stepanow}} \bibnamefont{and}
  \bibinfo{author}{\bibfnamefont{T.}~\bibnamefont{Thurn-Albrecht}},
  \bibinfo{journal}{Phys. Rev. E} \textbf{\bibinfo{volume}{79}},
  \bibinfo{pages}{041104} (\bibinfo{year}{2009}).

\bibitem[{\citenamefont{Wirtz and Fuller}(1993)}]{down2}
\bibinfo{author}{\bibfnamefont{D.}~\bibnamefont{Wirtz}} \bibnamefont{and}
  \bibinfo{author}{\bibfnamefont{G.~G.} \bibnamefont{Fuller}},
  \bibinfo{journal}{Phys. Rev. Lett.} \textbf{\bibinfo{volume}{71}},
  \bibinfo{pages}{2236} (\bibinfo{year}{1993}).

\bibitem[{\citenamefont{Maerzke and Siepmann}(2010)}]{SiepmannField}
\bibinfo{author}{\bibfnamefont{K.~A.} \bibnamefont{Maerzke}} \bibnamefont{and}
  \bibinfo{author}{\bibfnamefont{J.~I.} \bibnamefont{Siepmann}},
  \bibinfo{journal}{J. Phys. Chem. B} \textbf{\bibinfo{volume}{114}},
  \bibinfo{pages}{4261} (\bibinfo{year}{2010}).

\bibitem[{\citenamefont{Gao et~al.}(1999)\citenamefont{Gao, Oh, and
  C}}]{ZengField}
\bibinfo{author}{\bibfnamefont{G.~T.} \bibnamefont{Gao}},
  \bibinfo{author}{\bibfnamefont{K.~J.} \bibnamefont{Oh}}, \bibnamefont{and}
  \bibinfo{author}{\bibfnamefont{Z.~X.} \bibnamefont{C}}, \bibinfo{journal}{J.
  Chem. Phys.} \textbf{\bibinfo{volume}{110}}, \bibinfo{pages}{2533}
  (\bibinfo{year}{1999}).

\bibitem[{\citenamefont{Tsori and Leibler}(2007)}]{leibler}
\bibinfo{author}{\bibfnamefont{Y.}~\bibnamefont{Tsori}} \bibnamefont{and}
  \bibinfo{author}{\bibfnamefont{L.}~\bibnamefont{Leibler}},
  \bibinfo{journal}{Proc. Natl. Acad. Sci. U.S.A.}
  \textbf{\bibinfo{volume}{104}}, \bibinfo{pages}{7348} (\bibinfo{year}{2007}).

\bibitem[{\citenamefont{Gabor and Szalai}(2008)}]{Gabor}
\bibinfo{author}{\bibfnamefont{A.}~\bibnamefont{Gabor}} \bibnamefont{and}
  \bibinfo{author}{\bibfnamefont{I.}~\bibnamefont{Szalai}},
  \bibinfo{journal}{Mol. Phys.} \textbf{\bibinfo{volume}{106}},
  \bibinfo{pages}{801} (\bibinfo{year}{2008}).

\bibitem[{\citenamefont{Bhandary et~al.}(2014)\citenamefont{Bhandary,
  Srivastava, Srivastava, and Singh}}]{ConfSingh}
\bibinfo{author}{\bibfnamefont{D.}~\bibnamefont{Bhandary}},
  \bibinfo{author}{\bibfnamefont{K.}~\bibnamefont{Srivastava}},
  \bibinfo{author}{\bibfnamefont{R.}~\bibnamefont{Srivastava}},
  \bibnamefont{and} \bibinfo{author}{\bibfnamefont{J.~K.} \bibnamefont{Singh}},
  \bibinfo{journal}{J. Chem. Eng. Data} \textbf{\bibinfo{volume}{59}},
  \bibinfo{pages}{3090} (\bibinfo{year}{2014}).

\bibitem[{\citenamefont{Kessel et~al.}(2005)\citenamefont{Kessel, Ulmer,
  Pettke, Schmidt, and Thompson}}]{kessel2005water}
\bibinfo{author}{\bibfnamefont{R.}~\bibnamefont{Kessel}},
  \bibinfo{author}{\bibfnamefont{P.}~\bibnamefont{Ulmer}},
  \bibinfo{author}{\bibfnamefont{T.}~\bibnamefont{Pettke}},
  \bibinfo{author}{\bibfnamefont{M.}~\bibnamefont{Schmidt}}, \bibnamefont{and}
  \bibinfo{author}{\bibfnamefont{A.}~\bibnamefont{Thompson}},
  \bibinfo{journal}{Earth Planet. Sci. Lett.} \textbf{\bibinfo{volume}{237}},
  \bibinfo{pages}{873} (\bibinfo{year}{2005}).

\bibitem[{\citenamefont{Peterson et~al.}(2008)\citenamefont{Peterson, Vogel,
  Lachance, Fr{\"o}ling, Antal~Jr, and Tester}}]{peterson2008thermochemical}
\bibinfo{author}{\bibfnamefont{A.~A.} \bibnamefont{Peterson}},
  \bibinfo{author}{\bibfnamefont{F.}~\bibnamefont{Vogel}},
  \bibinfo{author}{\bibfnamefont{R.~P.} \bibnamefont{Lachance}},
  \bibinfo{author}{\bibfnamefont{M.}~\bibnamefont{Fr{\"o}ling}},
  \bibinfo{author}{\bibfnamefont{M.~J.} \bibnamefont{Antal~Jr}},
  \bibnamefont{and} \bibinfo{author}{\bibfnamefont{J.~W.}
  \bibnamefont{Tester}}, \bibinfo{journal}{Energy Environ. Sci.}
  \textbf{\bibinfo{volume}{1}}, \bibinfo{pages}{32} (\bibinfo{year}{2008}).

\bibitem[{\citenamefont{Desgranges and
  Delhommelle}(2012{\natexlab{a}})}]{PartI}
\bibinfo{author}{\bibfnamefont{C.}~\bibnamefont{Desgranges}} \bibnamefont{and}
  \bibinfo{author}{\bibfnamefont{J.}~\bibnamefont{Delhommelle}},
  \bibinfo{journal}{J. Chem. Phys.} \textbf{\bibinfo{volume}{136}},
  \bibinfo{pages}{184107} (\bibinfo{year}{2012}{\natexlab{a}}).

\bibitem[{\citenamefont{Desgranges and
  Delhommelle}(2012{\natexlab{b}})}]{PartII}
\bibinfo{author}{\bibfnamefont{C.}~\bibnamefont{Desgranges}} \bibnamefont{and}
  \bibinfo{author}{\bibfnamefont{J.}~\bibnamefont{Delhommelle}},
  \bibinfo{journal}{J. Chem. Phys.} \textbf{\bibinfo{volume}{136}},
  \bibinfo{pages}{184108} (\bibinfo{year}{2012}{\natexlab{b}}).

\bibitem[{\citenamefont{Desgranges and Delhommelle}(2014)}]{PartIII}
\bibinfo{author}{\bibfnamefont{C.}~\bibnamefont{Desgranges}} \bibnamefont{and}
  \bibinfo{author}{\bibfnamefont{J.}~\bibnamefont{Delhommelle}},
  \bibinfo{journal}{J. Chem. Phys.} \textbf{\bibinfo{volume}{140}},
  \bibinfo{pages}{104109} (\bibinfo{year}{2014}).

\bibitem[{\citenamefont{Desgranges and
  Delhommelle}(2016{\natexlab{a}})}]{PartIV}
\bibinfo{author}{\bibfnamefont{C.}~\bibnamefont{Desgranges}} \bibnamefont{and}
  \bibinfo{author}{\bibfnamefont{J.}~\bibnamefont{Delhommelle}},
  \bibinfo{journal}{J. Chem. Phys.} \textbf{\bibinfo{volume}{144}},
  \bibinfo{pages}{124510} (\bibinfo{year}{2016}{\natexlab{a}}).

\bibitem[{\citenamefont{Nedostup}(2013)}]{nedostup2013asymptotic}
\bibinfo{author}{\bibfnamefont{V.}~\bibnamefont{Nedostup}},
  \bibinfo{journal}{High Temperature} \textbf{\bibinfo{volume}{51}},
  \bibinfo{pages}{72} (\bibinfo{year}{2013}).

\bibitem[{\citenamefont{Apfelbaum and Vorob'ev}(2013)}]{apfelbaum2013regarding}
\bibinfo{author}{\bibfnamefont{E.}~\bibnamefont{Apfelbaum}} \bibnamefont{and}
  \bibinfo{author}{\bibfnamefont{V.}~\bibnamefont{Vorob'ev}},
  \bibinfo{journal}{J. Phys. Chem. B} \textbf{\bibinfo{volume}{117}},
  \bibinfo{pages}{7750} (\bibinfo{year}{2013}).

\bibitem[{\citenamefont{Kutney et~al.}(2000)\citenamefont{Kutney, Reagan,
  Smith, Tester, and Herschbach}}]{kutney2000zeno}
\bibinfo{author}{\bibfnamefont{M.~C.} \bibnamefont{Kutney}},
  \bibinfo{author}{\bibfnamefont{M.~T.} \bibnamefont{Reagan}},
  \bibinfo{author}{\bibfnamefont{K.~A.} \bibnamefont{Smith}},
  \bibinfo{author}{\bibfnamefont{J.~W.} \bibnamefont{Tester}},
  \bibnamefont{and} \bibinfo{author}{\bibfnamefont{D.~R.}
  \bibnamefont{Herschbach}}, \bibinfo{journal}{J. Phys. Chem. B}
  \textbf{\bibinfo{volume}{104}}, \bibinfo{pages}{9513} (\bibinfo{year}{2000}).

\bibitem[{\citenamefont{Wei and Herschbach}(2013)}]{wei2013isomorphism}
\bibinfo{author}{\bibfnamefont{Q.}~\bibnamefont{Wei}} \bibnamefont{and}
  \bibinfo{author}{\bibfnamefont{D.~R.} \bibnamefont{Herschbach}},
  \bibinfo{journal}{J. Phys. Chem. C} \textbf{\bibinfo{volume}{117}},
  \bibinfo{pages}{22438} (\bibinfo{year}{2013}).

\bibitem[{\citenamefont{Desgranges
  et~al.}(2016{\natexlab{a}})\citenamefont{Desgranges, Widhalm, and
  Delhommelle}}]{Leanna}
\bibinfo{author}{\bibfnamefont{C.}~\bibnamefont{Desgranges}},
  \bibinfo{author}{\bibfnamefont{L.}~\bibnamefont{Widhalm}}, \bibnamefont{and}
  \bibinfo{author}{\bibfnamefont{J.}~\bibnamefont{Delhommelle}},
  \bibinfo{journal}{J. Phys. Chem. B} \textbf{\bibinfo{volume}{120}},
  \bibinfo{pages}{5255} (\bibinfo{year}{2016}{\natexlab{a}}).

\bibitem[{\citenamefont{Desgranges and
  Delhommelle}(2016{\natexlab{b}})}]{Landon}
\bibinfo{author}{\bibfnamefont{L.}~\bibnamefont{Desgranges},
  \bibfnamefont{C~Huber}} \bibnamefont{and}
  \bibinfo{author}{\bibfnamefont{J.}~\bibnamefont{Delhommelle}},
  \bibinfo{journal}{Phys. Rev. E} \textbf{\bibinfo{volume}{94}},
  \bibinfo{pages}{012612} (\bibinfo{year}{2016}{\natexlab{b}}).

\bibitem[{\citenamefont{Apfelbaum and Vorob'ev}(2016)}]{Apfelbaum2016}
\bibinfo{author}{\bibfnamefont{E.}~\bibnamefont{Apfelbaum}} \bibnamefont{and}
  \bibinfo{author}{\bibfnamefont{V.}~\bibnamefont{Vorob'ev}},
  \bibinfo{journal}{J. Phys. Chem. B} \textbf{\bibinfo{volume}{120}},
  \bibinfo{pages}{4828} (\bibinfo{year}{2016}).

\bibitem[{\citenamefont{Brazhkin et~al.}(2013)\citenamefont{Brazhkin, Fomin,
  Lyapin, Ryzhov, Tsiok, and Trachenko}}]{brazhkin2013liquid}
\bibinfo{author}{\bibfnamefont{V.}~\bibnamefont{Brazhkin}},
  \bibinfo{author}{\bibfnamefont{Y.~D.} \bibnamefont{Fomin}},
  \bibinfo{author}{\bibfnamefont{A.}~\bibnamefont{Lyapin}},
  \bibinfo{author}{\bibfnamefont{V.}~\bibnamefont{Ryzhov}},
  \bibinfo{author}{\bibfnamefont{E.}~\bibnamefont{Tsiok}}, \bibnamefont{and}
  \bibinfo{author}{\bibfnamefont{K.}~\bibnamefont{Trachenko}},
  \bibinfo{journal}{Phys. Rev. Lett.} \textbf{\bibinfo{volume}{111}},
  \bibinfo{pages}{145901} (\bibinfo{year}{2013}).

\bibitem[{\citenamefont{Brazhkin and Ryzhov}(2011)}]{brazhkin2011van}
\bibinfo{author}{\bibfnamefont{V.}~\bibnamefont{Brazhkin}} \bibnamefont{and}
  \bibinfo{author}{\bibfnamefont{V.}~\bibnamefont{Ryzhov}},
  \bibinfo{journal}{J. Chem. Phys.} \textbf{\bibinfo{volume}{135}},
  \bibinfo{pages}{084503} (\bibinfo{year}{2011}).

\bibitem[{\citenamefont{Nishikawa et~al.}(2003)\citenamefont{Nishikawa, Kusano,
  Arai, and Morita}}]{Morita}
\bibinfo{author}{\bibfnamefont{K.}~\bibnamefont{Nishikawa}},
  \bibinfo{author}{\bibfnamefont{K.}~\bibnamefont{Kusano}},
  \bibinfo{author}{\bibfnamefont{A.~A.} \bibnamefont{Arai}}, \bibnamefont{and}
  \bibinfo{author}{\bibfnamefont{T.}~\bibnamefont{Morita}},
  \bibinfo{journal}{J. Chem. Phys.} \textbf{\bibinfo{volume}{118}},
  \bibinfo{pages}{1341} (\bibinfo{year}{2003}).

\bibitem[{\citenamefont{Brazhkin et~al.}(2012)\citenamefont{Brazhkin, Lyapin,
  Ryzhov, Trachenko, Fomin, and Tsiok}}]{brazhkin2012supercritical}
\bibinfo{author}{\bibfnamefont{V.~V.} \bibnamefont{Brazhkin}},
  \bibinfo{author}{\bibfnamefont{A.~G.} \bibnamefont{Lyapin}},
  \bibinfo{author}{\bibfnamefont{V.~N.} \bibnamefont{Ryzhov}},
  \bibinfo{author}{\bibfnamefont{K.}~\bibnamefont{Trachenko}},
  \bibinfo{author}{\bibfnamefont{Y.~D.} \bibnamefont{Fomin}}, \bibnamefont{and}
  \bibinfo{author}{\bibfnamefont{E.~N.} \bibnamefont{Tsiok}},
  \bibinfo{journal}{Physics-Uspekhi} \textbf{\bibinfo{volume}{55}},
  \bibinfo{pages}{1061} (\bibinfo{year}{2012}).

\bibitem[{\citenamefont{Apfelbaum and
  Vorob'ev}(2015)}]{apfelbaum2015similarity}
\bibinfo{author}{\bibfnamefont{E.}~\bibnamefont{Apfelbaum}} \bibnamefont{and}
  \bibinfo{author}{\bibfnamefont{V.}~\bibnamefont{Vorob'ev}},
  \bibinfo{journal}{J. Phys. Chem. B} \textbf{\bibinfo{volume}{119}},
  \bibinfo{pages}{8419} (\bibinfo{year}{2015}).

\bibitem[{\citenamefont{Desgranges
  et~al.}(2016{\natexlab{b}})\citenamefont{Desgranges, Margo, and
  Delhommelle}}]{Abigail}
\bibinfo{author}{\bibfnamefont{C.}~\bibnamefont{Desgranges}},
  \bibinfo{author}{\bibfnamefont{A.}~\bibnamefont{Margo}}, \bibnamefont{and}
  \bibinfo{author}{\bibfnamefont{J.}~\bibnamefont{Delhommelle}},
  \bibinfo{journal}{Chem. Phys. Lett.} \textbf{\bibinfo{volume}{658}},
  \bibinfo{pages}{37} (\bibinfo{year}{2016}{\natexlab{b}}).

\bibitem[{\citenamefont{McQuarrie}(1976)}]{McQuarrie}
\bibinfo{author}{\bibfnamefont{D.~A.} \bibnamefont{McQuarrie}},
  \emph{\bibinfo{title}{Statistical Mechanics}} (\bibinfo{publisher}{Harper \&
  Row, New York}, \bibinfo{year}{1976}).

\bibitem[{\citenamefont{Lyubartsev et~al.}(1992)\citenamefont{Lyubartsev,
  Martsinovski, Shevkunov, and Vorontsov-Velyaminov}}]{Lyubartsev}
\bibinfo{author}{\bibfnamefont{A.~P.} \bibnamefont{Lyubartsev}},
  \bibinfo{author}{\bibfnamefont{A.~A.} \bibnamefont{Martsinovski}},
  \bibinfo{author}{\bibfnamefont{S.~V.} \bibnamefont{Shevkunov}},
  \bibnamefont{and} \bibinfo{author}{\bibfnamefont{P.~N.}
  \bibnamefont{Vorontsov-Velyaminov}}, \bibinfo{journal}{J. Chem. Phys.}
  \textbf{\bibinfo{volume}{96}}, \bibinfo{pages}{1776} (\bibinfo{year}{1992}).

\bibitem[{\citenamefont{Escobedo and de~Pablo}(1996)}]{expanded}
\bibinfo{author}{\bibfnamefont{F.}~\bibnamefont{Escobedo}} \bibnamefont{and}
  \bibinfo{author}{\bibfnamefont{J.~J.} \bibnamefont{de~Pablo}},
  \bibinfo{journal}{J. Chem. Phys.} \textbf{\bibinfo{volume}{105}},
  \bibinfo{pages}{4391} (\bibinfo{year}{1996}).

\bibitem[{\citenamefont{Muller and Paul}(1994)}]{Paul}
\bibinfo{author}{\bibfnamefont{M.}~\bibnamefont{Muller}} \bibnamefont{and}
  \bibinfo{author}{\bibfnamefont{W.}~\bibnamefont{Paul}}, \bibinfo{journal}{J.
  Chem. Phys.} \textbf{\bibinfo{volume}{100}}, \bibinfo{pages}{719}
  (\bibinfo{year}{1994}).

\bibitem[{\citenamefont{Escobedo and Abreu}(2006)}]{Escobedo}
\bibinfo{author}{\bibfnamefont{F.~A.} \bibnamefont{Escobedo}} \bibnamefont{and}
  \bibinfo{author}{\bibfnamefont{C.~R.~A.} \bibnamefont{Abreu}},
  \bibinfo{journal}{J. Chem. Phys.} \textbf{\bibinfo{volume}{124}},
  \bibinfo{pages}{104110} (\bibinfo{year}{2006}).

\bibitem[{\citenamefont{Singh and Errington}(2006)}]{Singh}
\bibinfo{author}{\bibfnamefont{J.~K.} \bibnamefont{Singh}} \bibnamefont{and}
  \bibinfo{author}{\bibfnamefont{J.~R.} \bibnamefont{Errington}},
  \bibinfo{journal}{J. Phys. Chem. B} \textbf{\bibinfo{volume}{110}},
  \bibinfo{pages}{1369} (\bibinfo{year}{2006}).

\bibitem[{\citenamefont{Escobedo and Martinez-Veracoechea}(2008)}]{MV2}
\bibinfo{author}{\bibfnamefont{F.~A.} \bibnamefont{Escobedo}} \bibnamefont{and}
  \bibinfo{author}{\bibfnamefont{F.~J.} \bibnamefont{Martinez-Veracoechea}},
  \bibinfo{journal}{J. Chem. Phys.} \textbf{\bibinfo{volume}{129}},
  \bibinfo{pages}{154107} (\bibinfo{year}{2008}).

\bibitem[{\citenamefont{Shi and Maginn}(2007)}]{Shi}
\bibinfo{author}{\bibfnamefont{W.}~\bibnamefont{Shi}} \bibnamefont{and}
  \bibinfo{author}{\bibfnamefont{E.~J.} \bibnamefont{Maginn}},
  \bibinfo{journal}{J. Chem. Theory Comp.} \textbf{\bibinfo{volume}{3}},
  \bibinfo{pages}{1451} (\bibinfo{year}{2007}).

\bibitem[{\citenamefont{Hicks et~al.}(2012)\citenamefont{Hicks, Desgranges, and
  Delhommelle}}]{Jason}
\bibinfo{author}{\bibfnamefont{J.~M.} \bibnamefont{Hicks}},
  \bibinfo{author}{\bibfnamefont{C.}~\bibnamefont{Desgranges}},
  \bibnamefont{and}
  \bibinfo{author}{\bibfnamefont{J.}~\bibnamefont{Delhommelle}},
  \bibinfo{journal}{J. Phys. Chem. C} \textbf{\bibinfo{volume}{116}},
  \bibinfo{pages}{22938} (\bibinfo{year}{2012}).

\bibitem[{\citenamefont{Koenig et~al.}(2014)\citenamefont{Koenig, Desgranges,
  and Delhommelle}}]{Aaron}
\bibinfo{author}{\bibfnamefont{A.~R.~V.} \bibnamefont{Koenig}},
  \bibinfo{author}{\bibfnamefont{C.}~\bibnamefont{Desgranges}},
  \bibnamefont{and}
  \bibinfo{author}{\bibfnamefont{J.}~\bibnamefont{Delhommelle}},
  \bibinfo{journal}{Molec. Simul.} \textbf{\bibinfo{volume}{40}},
  \bibinfo{pages}{71} (\bibinfo{year}{2014}).

\bibitem[{\citenamefont{Hicks et~al.}(2014)\citenamefont{Hicks, Desgranges, and
  Delhommelle}}]{Erica}
\bibinfo{author}{\bibfnamefont{E.~A.} \bibnamefont{Hicks}},
  \bibinfo{author}{\bibfnamefont{C.}~\bibnamefont{Desgranges}},
  \bibnamefont{and}
  \bibinfo{author}{\bibfnamefont{J.}~\bibnamefont{Delhommelle}},
  \bibinfo{journal}{Molec. Simul.} \textbf{\bibinfo{volume}{40}},
  \bibinfo{pages}{656} (\bibinfo{year}{2014}).

\bibitem[{\citenamefont{Owen et~al.}(2015)\citenamefont{Owen, Desgranges, and
  Delhommelle}}]{Andrew}
\bibinfo{author}{\bibfnamefont{A.~N.} \bibnamefont{Owen}},
  \bibinfo{author}{\bibfnamefont{C.}~\bibnamefont{Desgranges}},
  \bibnamefont{and}
  \bibinfo{author}{\bibfnamefont{J.}~\bibnamefont{Delhommelle}},
  \bibinfo{journal}{Fluid Phase Equil.} \textbf{\bibinfo{volume}{402}},
  \bibinfo{pages}{69} (\bibinfo{year}{2015}).

\bibitem[{\citenamefont{Rane et~al.}(2013)\citenamefont{Rane, Murali, and
  Errington}}]{Rane}
\bibinfo{author}{\bibfnamefont{K.~S.} \bibnamefont{Rane}},
  \bibinfo{author}{\bibfnamefont{S.}~\bibnamefont{Murali}}, \bibnamefont{and}
  \bibinfo{author}{\bibfnamefont{J.~R.} \bibnamefont{Errington}},
  \bibinfo{journal}{J. Chem. Theory Comput.} \textbf{\bibinfo{volume}{9}},
  \bibinfo{pages}{2552} (\bibinfo{year}{2013}).

\bibitem[{\citenamefont{Yee et~al.}(2013)\citenamefont{Yee, Shah, and
  Maginn}}]{Yee}
\bibinfo{author}{\bibfnamefont{P.}~\bibnamefont{Yee}},
  \bibinfo{author}{\bibfnamefont{J.~K.} \bibnamefont{Shah}}, \bibnamefont{and}
  \bibinfo{author}{\bibfnamefont{E.~J.} \bibnamefont{Maginn}},
  \bibinfo{journal}{J. Phys. Chem. B} \textbf{\bibinfo{volume}{117}},
  \bibinfo{pages}{12556} (\bibinfo{year}{2013}).

\bibitem[{\citenamefont{Sikora et~al.}(2015)\citenamefont{Sikora, Col{\`o}n,
  and Snurr}}]{Sikora}
\bibinfo{author}{\bibfnamefont{B.~J.} \bibnamefont{Sikora}},
  \bibinfo{author}{\bibfnamefont{Y.~J.} \bibnamefont{Col{\`o}n}},
  \bibnamefont{and} \bibinfo{author}{\bibfnamefont{R.~Q.} \bibnamefont{Snurr}},
  \bibinfo{journal}{Molec. Simul.} \textbf{\bibinfo{volume}{41}},
  \bibinfo{pages}{1339} (\bibinfo{year}{2015}).

\bibitem[{\citenamefont{Gazenm$\ddot{\mathrm{u}}$ller and Camp}(2007)}]{Camp}
\bibinfo{author}{\bibfnamefont{G.}~\bibnamefont{Gazenm$\ddot{\mathrm{u}}$ller}}
  \bibnamefont{and} \bibinfo{author}{\bibfnamefont{P.~J.} \bibnamefont{Camp}},
  \bibinfo{journal}{J. Chem. Phys.} \textbf{\bibinfo{volume}{127}},
  \bibinfo{pages}{154504} (\bibinfo{year}{2007}).

\bibitem[{\citenamefont{Wang and Landau}(2001{\natexlab{a}})}]{Wang1}
\bibinfo{author}{\bibfnamefont{F.}~\bibnamefont{Wang}} \bibnamefont{and}
  \bibinfo{author}{\bibfnamefont{D.~P.} \bibnamefont{Landau}},
  \bibinfo{journal}{Phys. Rev. E} \textbf{\bibinfo{volume}{64}},
  \bibinfo{pages}{056101} (\bibinfo{year}{2001}{\natexlab{a}}).

\bibitem[{\citenamefont{Wang and Landau}(2001{\natexlab{b}})}]{Wang2}
\bibinfo{author}{\bibfnamefont{F.}~\bibnamefont{Wang}} \bibnamefont{and}
  \bibinfo{author}{\bibfnamefont{D.}~\bibnamefont{Landau}},
  \bibinfo{journal}{Phys. Rev. Lett.} \textbf{\bibinfo{volume}{86}},
  \bibinfo{pages}{2050} (\bibinfo{year}{2001}{\natexlab{b}}).

\bibitem[{\citenamefont{Shell et~al.}(2002)\citenamefont{Shell, Debenedetti,
  and Panagiotopoulos}}]{Shell}
\bibinfo{author}{\bibfnamefont{M.~S.} \bibnamefont{Shell}},
  \bibinfo{author}{\bibfnamefont{P.~G.} \bibnamefont{Debenedetti}},
  \bibnamefont{and} \bibinfo{author}{\bibfnamefont{A.~Z.}
  \bibnamefont{Panagiotopoulos}}, \bibinfo{journal}{Phys. Rev. E}
  \textbf{\bibinfo{volume}{66}}, \bibinfo{pages}{056703}
  (\bibinfo{year}{2002}).

\bibitem[{\citenamefont{Yan et~al.}(2002)\citenamefont{Yan, Faller, and
  de~Pablo}}]{dePablo}
\bibinfo{author}{\bibfnamefont{Q.}~\bibnamefont{Yan}},
  \bibinfo{author}{\bibfnamefont{R.}~\bibnamefont{Faller}}, \bibnamefont{and}
  \bibinfo{author}{\bibfnamefont{J.~J.} \bibnamefont{de~Pablo}},
  \bibinfo{journal}{J. Chem. Phys.} \textbf{\bibinfo{volume}{116}},
  \bibinfo{pages}{8745} (\bibinfo{year}{2002}).

\bibitem[{\citenamefont{Luettmer-Strathmann
  et~al.}(2008)\citenamefont{Luettmer-Strathmann, Rampf, Paul, and
  Binder}}]{Rampf}
\bibinfo{author}{\bibfnamefont{J.}~\bibnamefont{Luettmer-Strathmann}},
  \bibinfo{author}{\bibfnamefont{F.}~\bibnamefont{Rampf}},
  \bibinfo{author}{\bibfnamefont{W.}~\bibnamefont{Paul}}, \bibnamefont{and}
  \bibinfo{author}{\bibfnamefont{K.}~\bibnamefont{Binder}},
  \bibinfo{journal}{J. Chem. Phys.} \textbf{\bibinfo{volume}{128}},
  \bibinfo{pages}{064903} (\bibinfo{year}{2008}).

\bibitem[{\citenamefont{Aleksandrov et~al.}(2010)\citenamefont{Aleksandrov,
  Desgranges, and Delhommelle}}]{Copper}
\bibinfo{author}{\bibfnamefont{T.}~\bibnamefont{Aleksandrov}},
  \bibinfo{author}{\bibfnamefont{C.}~\bibnamefont{Desgranges}},
  \bibnamefont{and}
  \bibinfo{author}{\bibfnamefont{J.}~\bibnamefont{Delhommelle}},
  \bibinfo{journal}{Fluid Phase Equil.} \textbf{\bibinfo{volume}{287}},
  \bibinfo{pages}{79} (\bibinfo{year}{2010}).

\bibitem[{\citenamefont{Desgranges et~al.}(2010)\citenamefont{Desgranges,
  Hicks, Magness, and Delhommelle}}]{MolPhys}
\bibinfo{author}{\bibfnamefont{C.}~\bibnamefont{Desgranges}},
  \bibinfo{author}{\bibfnamefont{J.~M.} \bibnamefont{Hicks}},
  \bibinfo{author}{\bibfnamefont{A.}~\bibnamefont{Magness}}, \bibnamefont{and}
  \bibinfo{author}{\bibfnamefont{J.}~\bibnamefont{Delhommelle}},
  \bibinfo{journal}{Mol. Phys.} \textbf{\bibinfo{volume}{108}},
  \bibinfo{pages}{151} (\bibinfo{year}{2010}).

\bibitem[{\citenamefont{Ngale et~al.}(2012)\citenamefont{Ngale, Desgranges, and
  Delhommelle}}]{KennethI}
\bibinfo{author}{\bibfnamefont{K.~N.} \bibnamefont{Ngale}},
  \bibinfo{author}{\bibfnamefont{C.}~\bibnamefont{Desgranges}},
  \bibnamefont{and}
  \bibinfo{author}{\bibfnamefont{J.}~\bibnamefont{Delhommelle}},
  \bibinfo{journal}{Molec. Simul.} \textbf{\bibinfo{volume}{38}},
  \bibinfo{pages}{653} (\bibinfo{year}{2012}).

\bibitem[{\citenamefont{Shell et~al.}(2003)\citenamefont{Shell, Debenedetti,
  and Panagiotopoulos}}]{Shell2}
\bibinfo{author}{\bibfnamefont{M.~S.} \bibnamefont{Shell}},
  \bibinfo{author}{\bibfnamefont{P.~G.} \bibnamefont{Debenedetti}},
  \bibnamefont{and} \bibinfo{author}{\bibfnamefont{A.~Z.}
  \bibnamefont{Panagiotopoulos}}, \bibinfo{journal}{J. Chem. Phys.}
  \textbf{\bibinfo{volume}{119}}, \bibinfo{pages}{9406} (\bibinfo{year}{2003}).

\bibitem[{\citenamefont{Shell et~al.}(2004)\citenamefont{Shell, Debenedetti,
  and Panagiotopoulos}}]{Shell3}
\bibinfo{author}{\bibfnamefont{M.~S.} \bibnamefont{Shell}},
  \bibinfo{author}{\bibfnamefont{P.~G.} \bibnamefont{Debenedetti}},
  \bibnamefont{and} \bibinfo{author}{\bibfnamefont{A.~Z.}
  \bibnamefont{Panagiotopoulos}}, \bibinfo{journal}{J. Phys. Chem. B}
  \textbf{\bibinfo{volume}{108}}, \bibinfo{pages}{19748}
  (\bibinfo{year}{2004}).

\bibitem[{\citenamefont{Malakis et~al.}(2010)\citenamefont{Malakis, Berker,
  Hijagapiou, Fytas, and Papakonstantinou}}]{Malakis}
\bibinfo{author}{\bibfnamefont{A.}~\bibnamefont{Malakis}},
  \bibinfo{author}{\bibfnamefont{A.~N.} \bibnamefont{Berker}},
  \bibinfo{author}{\bibfnamefont{I.~A.} \bibnamefont{Hijagapiou}},
  \bibinfo{author}{\bibfnamefont{N.~G.} \bibnamefont{Fytas}}, \bibnamefont{and}
  \bibinfo{author}{\bibfnamefont{T.}~\bibnamefont{Papakonstantinou}},
  \bibinfo{journal}{Phys. Rev. E} \textbf{\bibinfo{volume}{81}},
  \bibinfo{pages}{041113} (\bibinfo{year}{2010}).

\bibitem[{\citenamefont{Desgranges and Delhommelle}(2009)}]{WLHMC}
\bibinfo{author}{\bibfnamefont{C.}~\bibnamefont{Desgranges}} \bibnamefont{and}
  \bibinfo{author}{\bibfnamefont{J.}~\bibnamefont{Delhommelle}},
  \bibinfo{journal}{J. Chem. Phys.} \textbf{\bibinfo{volume}{130}},
  \bibinfo{pages}{244109} (\bibinfo{year}{2009}).

\bibitem[{\citenamefont{Do et~al.}(2011)\citenamefont{Do, Hirst, and
  Wheatley}}]{Do}
\bibinfo{author}{\bibfnamefont{H.}~\bibnamefont{Do}},
  \bibinfo{author}{\bibfnamefont{J.}~\bibnamefont{Hirst}}, \bibnamefont{and}
  \bibinfo{author}{\bibfnamefont{R.}~\bibnamefont{Wheatley}},
  \bibinfo{journal}{J. Chem. Phys.} \textbf{\bibinfo{volume}{135}},
  \bibinfo{pages}{174105} (\bibinfo{year}{2011}).

\bibitem[{\citenamefont{Berendsen et~al.}(1987)\citenamefont{Berendsen,
  Grigera, and Straatsma}}]{berendsen1987missing}
\bibinfo{author}{\bibfnamefont{H.}~\bibnamefont{Berendsen}},
  \bibinfo{author}{\bibfnamefont{J.}~\bibnamefont{Grigera}}, \bibnamefont{and}
  \bibinfo{author}{\bibfnamefont{T.}~\bibnamefont{Straatsma}},
  \bibinfo{journal}{J. Phys. Chem.} \textbf{\bibinfo{volume}{91}},
  \bibinfo{pages}{6269} (\bibinfo{year}{1987}).

\bibitem[{\citenamefont{Aragones et~al.}(2011)\citenamefont{Aragones,
  MacDowell, Siepmann, and Vega}}]{aragones2011phase}
\bibinfo{author}{\bibfnamefont{J.}~\bibnamefont{Aragones}},
  \bibinfo{author}{\bibfnamefont{L.}~\bibnamefont{MacDowell}},
  \bibinfo{author}{\bibfnamefont{J.}~\bibnamefont{Siepmann}}, \bibnamefont{and}
  \bibinfo{author}{\bibfnamefont{C.}~\bibnamefont{Vega}},
  \bibinfo{journal}{Phys. Rev. Lett.} \textbf{\bibinfo{volume}{107}},
  \bibinfo{pages}{155702} (\bibinfo{year}{2011}).

\bibitem[{\citenamefont{Saitta et~al.}(2012)\citenamefont{Saitta, Saija, and
  Giaquinta}}]{saitta2012ab}
\bibinfo{author}{\bibfnamefont{A.~M.} \bibnamefont{Saitta}},
  \bibinfo{author}{\bibfnamefont{F.}~\bibnamefont{Saija}}, \bibnamefont{and}
  \bibinfo{author}{\bibfnamefont{P.~V.} \bibnamefont{Giaquinta}},
  \bibinfo{journal}{Phys. Rev. Lett.} \textbf{\bibinfo{volume}{108}},
  \bibinfo{pages}{207801} (\bibinfo{year}{2012}).

\bibitem[{\citenamefont{Stuve}(2012)}]{stuve2012ionization}
\bibinfo{author}{\bibfnamefont{E.~M.} \bibnamefont{Stuve}},
  \bibinfo{journal}{Chem. Phys. Lett.} \textbf{\bibinfo{volume}{519}},
  \bibinfo{pages}{1} (\bibinfo{year}{2012}).

\bibitem[{\citenamefont{Rothfuss et~al.}(2003)\citenamefont{Rothfuss, Medvedev,
  and Stuve}}]{rothfuss2003influence}
\bibinfo{author}{\bibfnamefont{C.~J.} \bibnamefont{Rothfuss}},
  \bibinfo{author}{\bibfnamefont{V.~K.} \bibnamefont{Medvedev}},
  \bibnamefont{and} \bibinfo{author}{\bibfnamefont{E.~M.} \bibnamefont{Stuve}},
  \bibinfo{journal}{J. Electroanal. Chem.} \textbf{\bibinfo{volume}{554}},
  \bibinfo{pages}{133} (\bibinfo{year}{2003}).

\bibitem[{\citenamefont{Geissler et~al.}(2001)\citenamefont{Geissler, Dellago,
  Chandler, Hutter, and Parrinello}}]{geissler2001autoionization}
\bibinfo{author}{\bibfnamefont{P.~L.} \bibnamefont{Geissler}},
  \bibinfo{author}{\bibfnamefont{C.}~\bibnamefont{Dellago}},
  \bibinfo{author}{\bibfnamefont{D.}~\bibnamefont{Chandler}},
  \bibinfo{author}{\bibfnamefont{J.}~\bibnamefont{Hutter}}, \bibnamefont{and}
  \bibinfo{author}{\bibfnamefont{M.}~\bibnamefont{Parrinello}},
  \bibinfo{journal}{Science} \textbf{\bibinfo{volume}{291}},
  \bibinfo{pages}{2121} (\bibinfo{year}{2001}).

\bibitem[{\citenamefont{Allen and Tildesley}(1987)}]{Allen}
\bibinfo{author}{\bibfnamefont{M.~P.} \bibnamefont{Allen}} \bibnamefont{and}
  \bibinfo{author}{\bibfnamefont{D.~J.} \bibnamefont{Tildesley}},
  \emph{\bibinfo{title}{Computer Simulation of Liquids}}
  (\bibinfo{publisher}{Clarendon Press, Oxford}, \bibinfo{year}{1987}).

\bibitem[{\citenamefont{Desgranges and Delhommelle}(2015)}]{jctc2015}
\bibinfo{author}{\bibfnamefont{C.}~\bibnamefont{Desgranges}} \bibnamefont{and}
  \bibinfo{author}{\bibfnamefont{J.}~\bibnamefont{Delhommelle}},
  \bibinfo{journal}{J. Chem. Theory Comput.} \textbf{\bibinfo{volume}{11}},
  \bibinfo{pages}{5401} (\bibinfo{year}{2015}).

\bibitem[{\citenamefont{Kiselev and Heinzinger}(1996)}]{gr1}
\bibinfo{author}{\bibfnamefont{M.}~\bibnamefont{Kiselev}} \bibnamefont{and}
  \bibinfo{author}{\bibfnamefont{K.}~\bibnamefont{Heinzinger}},
  \bibinfo{journal}{J. Chem. Phys.} \textbf{\bibinfo{volume}{105}},
  \bibinfo{pages}{650} (\bibinfo{year}{1996}).

\bibitem[{\citenamefont{Jung et~al.}(1999)\citenamefont{Jung, Yang, and
  Jhon}}]{gr2}
\bibinfo{author}{\bibfnamefont{D.~H.} \bibnamefont{Jung}},
  \bibinfo{author}{\bibfnamefont{J.~H.} \bibnamefont{Yang}}, \bibnamefont{and}
  \bibinfo{author}{\bibfnamefont{M.~S.} \bibnamefont{Jhon}},
  \bibinfo{journal}{Chem. Phys.} \textbf{\bibinfo{volume}{244}},
  \bibinfo{pages}{331} (\bibinfo{year}{1999}).

\bibitem[{\citenamefont{Sun et~al.}(2005{\natexlab{a}})\citenamefont{Sun, Chen,
  and Huang}}]{gr3}
\bibinfo{author}{\bibfnamefont{W.}~\bibnamefont{Sun}},
  \bibinfo{author}{\bibfnamefont{Z.}~\bibnamefont{Chen}}, \bibnamefont{and}
  \bibinfo{author}{\bibfnamefont{S.-Y.} \bibnamefont{Huang}},
  \bibinfo{journal}{Molec. Simul.} \textbf{\bibinfo{volume}{31}},
  \bibinfo{pages}{555} (\bibinfo{year}{2005}{\natexlab{a}}).

\bibitem[{\citenamefont{Sun et~al.}(2005{\natexlab{b}})\citenamefont{Sun, Chen,
  and Huang}}]{gr4}
\bibinfo{author}{\bibfnamefont{W.}~\bibnamefont{Sun}},
  \bibinfo{author}{\bibfnamefont{Z.}~\bibnamefont{Chen}}, \bibnamefont{and}
  \bibinfo{author}{\bibfnamefont{S.-Y.} \bibnamefont{Huang}},
  \bibinfo{journal}{Fluid Phase Equil.} \textbf{\bibinfo{volume}{238}},
  \bibinfo{pages}{20} (\bibinfo{year}{2005}{\natexlab{b}}).

\bibitem[{\citenamefont{Evans and Morriss}(1990)}]{theBook}
\bibinfo{author}{\bibfnamefont{D.~J.} \bibnamefont{Evans}} \bibnamefont{and}
  \bibinfo{author}{\bibfnamefont{G.~P.} \bibnamefont{Morriss}},
  \emph{\bibinfo{title}{Statistical Mechanics of Nonequilibrium Liquids}}
  (\bibinfo{publisher}{Academic Press, London}, \bibinfo{year}{1990}).

\bibitem[{\citenamefont{Petravic and Delhommelle}(2005)}]{EthanolShear}
\bibinfo{author}{\bibfnamefont{J.}~\bibnamefont{Petravic}} \bibnamefont{and}
  \bibinfo{author}{\bibfnamefont{J.}~\bibnamefont{Delhommelle}},
  \bibinfo{journal}{J. Chem. Phys.} \textbf{\bibinfo{volume}{122}},
  \bibinfo{pages}{234509} (\bibinfo{year}{2005}).

\bibitem[{\citenamefont{Chau and Hardwick}(1998)}]{Chau}
\bibinfo{author}{\bibfnamefont{P.-L.} \bibnamefont{Chau}} \bibnamefont{and}
  \bibinfo{author}{\bibfnamefont{A.~J.} \bibnamefont{Hardwick}},
  \bibinfo{journal}{Mol. Phys.} \textbf{\bibinfo{volume}{93}},
  \bibinfo{pages}{511} (\bibinfo{year}{1998}).

\bibitem[{\citenamefont{Errington and Debenedetti}(2001)}]{pdebene}
\bibinfo{author}{\bibfnamefont{J.~R.} \bibnamefont{Errington}}
  \bibnamefont{and} \bibinfo{author}{\bibfnamefont{P.~G.}
  \bibnamefont{Debenedetti}}, \bibinfo{journal}{Nature}
  \textbf{\bibinfo{volume}{409}}, \bibinfo{pages}{318} (\bibinfo{year}{2001}).

\bibitem[{\citenamefont{Zielkiewicz}(2005)}]{Hbond1}
\bibinfo{author}{\bibfnamefont{J.}~\bibnamefont{Zielkiewicz}},
  \bibinfo{journal}{J. Chem. Phys.} \textbf{\bibinfo{volume}{123}},
  \bibinfo{pages}{104501} (\bibinfo{year}{2005}).

\bibitem[{\citenamefont{Luzar and D}(1996{\natexlab{a}})}]{Hbond2}
\bibinfo{author}{\bibfnamefont{A.}~\bibnamefont{Luzar}} \bibnamefont{and}
  \bibinfo{author}{\bibfnamefont{C.}~\bibnamefont{D}}, \bibinfo{journal}{Phys.
  Rev. Lett.} \textbf{\bibinfo{volume}{76}}, \bibinfo{pages}{928}
  (\bibinfo{year}{1996}{\natexlab{a}}).

\bibitem[{\citenamefont{Luzar and D}(1996{\natexlab{b}})}]{Hbond3}
\bibinfo{author}{\bibfnamefont{A.}~\bibnamefont{Luzar}} \bibnamefont{and}
  \bibinfo{author}{\bibfnamefont{C.}~\bibnamefont{D}},
  \bibinfo{journal}{Nature} \textbf{\bibinfo{volume}{379}}, \bibinfo{pages}{55}
  (\bibinfo{year}{1996}{\natexlab{b}}).

\bibitem[{\citenamefont{Zielkiewicz}(2000)}]{Hbond4}
\bibinfo{author}{\bibfnamefont{J.}~\bibnamefont{Zielkiewicz}},
  \bibinfo{journal}{Phys. Rev. E} \textbf{\bibinfo{volume}{62}},
  \bibinfo{pages}{579} (\bibinfo{year}{2000}).

\bibitem[{\citenamefont{Apfelbaum and
  Vorob'ev}(2009{\natexlab{a}})}]{apfelbaum2009confirmation}
\bibinfo{author}{\bibfnamefont{E.}~\bibnamefont{Apfelbaum}} \bibnamefont{and}
  \bibinfo{author}{\bibfnamefont{V.}~\bibnamefont{Vorob'ev}},
  \bibinfo{journal}{J. Chem. Phys.} \textbf{\bibinfo{volume}{130}},
  \bibinfo{pages}{214111} (\bibinfo{year}{2009}{\natexlab{a}}).

\bibitem[{\citenamefont{Apfelbaum and
  Vorob'ev}(2009{\natexlab{b}})}]{apfelbaum2009correspondence}
\bibinfo{author}{\bibfnamefont{E.}~\bibnamefont{Apfelbaum}} \bibnamefont{and}
  \bibinfo{author}{\bibfnamefont{V.}~\bibnamefont{Vorob'ev}},
  \bibinfo{journal}{J. Phys. Chem. B} \textbf{\bibinfo{volume}{113}},
  \bibinfo{pages}{3521} (\bibinfo{year}{2009}{\natexlab{b}}).

\bibitem[{\citenamefont{Apfelbaum and Vorob'ev}(2008)}]{apfelbaum2008new}
\bibinfo{author}{\bibfnamefont{E.}~\bibnamefont{Apfelbaum}} \bibnamefont{and}
  \bibinfo{author}{\bibfnamefont{V.}~\bibnamefont{Vorob'ev}},
  \bibinfo{journal}{J. Phys. Chem. B} \textbf{\bibinfo{volume}{112}},
  \bibinfo{pages}{13064} (\bibinfo{year}{2008}).

\bibitem[{\citenamefont{Ceriotti et~al.}(2016)\citenamefont{Ceriotti, Fang,
  Kusalik, McKenzie, Michaelides, Morales, and Markland}}]{ceriotti2016nuclear}
\bibinfo{author}{\bibfnamefont{M.}~\bibnamefont{Ceriotti}},
  \bibinfo{author}{\bibfnamefont{W.}~\bibnamefont{Fang}},
  \bibinfo{author}{\bibfnamefont{P.~G.} \bibnamefont{Kusalik}},
  \bibinfo{author}{\bibfnamefont{R.~H.} \bibnamefont{McKenzie}},
  \bibinfo{author}{\bibfnamefont{A.}~\bibnamefont{Michaelides}},
  \bibinfo{author}{\bibfnamefont{M.~A.} \bibnamefont{Morales}},
  \bibnamefont{and} \bibinfo{author}{\bibfnamefont{T.~E.}
  \bibnamefont{Markland}}, \bibinfo{journal}{Chem. Rev.}
  \textbf{\bibinfo{volume}{116}}, \bibinfo{pages}{7529} (\bibinfo{year}{2016}).

\bibitem[{\citenamefont{Chau et~al.}(2001)\citenamefont{Chau, Mitchell, Minich,
  and Nellis}}]{chau2001electrical}
\bibinfo{author}{\bibfnamefont{R.}~\bibnamefont{Chau}},
  \bibinfo{author}{\bibfnamefont{A.}~\bibnamefont{Mitchell}},
  \bibinfo{author}{\bibfnamefont{R.}~\bibnamefont{Minich}}, \bibnamefont{and}
  \bibinfo{author}{\bibfnamefont{W.}~\bibnamefont{Nellis}},
  \bibinfo{journal}{J. Chem. Phys.} \textbf{\bibinfo{volume}{114}},
  \bibinfo{pages}{1361} (\bibinfo{year}{2001}).

\bibitem[{\citenamefont{Chau et~al.}(2011)\citenamefont{Chau, Hamel, and
  Nellis}}]{chau2011chemical}
\bibinfo{author}{\bibfnamefont{R.}~\bibnamefont{Chau}},
  \bibinfo{author}{\bibfnamefont{S.}~\bibnamefont{Hamel}}, \bibnamefont{and}
  \bibinfo{author}{\bibfnamefont{W.~J.} \bibnamefont{Nellis}},
  \bibinfo{journal}{Nature Commun.} \textbf{\bibinfo{volume}{2}},
  \bibinfo{pages}{203} (\bibinfo{year}{2011}).

\end{thebibliography}
\end{document}